\documentclass[11pt,letterpaper]{article}
\usepackage{tabularx,environ,amsmath,amssymb,amsthm}
\usepackage{tikz}
\usepackage{mathtools} 
\usepackage{amsthm,amssymb,amsmath,mathrsfs,bbm,stmaryrd}
\usepackage[top=1in,right=1in,bottom=1in,left=1in]{geometry}      
\usepackage[linesnumbered,inoutnumbered]{algorithm2e}
\usepackage{graphics}
\usepackage{comment}
\usepackage{xcolor}
\usepackage{lineno}
\usepackage{setspace}
\usepackage{multirow}
\usepackage{tablefootnote}
\usepackage{longtable}
\usepackage{hyperref} 

\usetikzlibrary{arrows}
\newcommand{\mypsi}[2]{\ensuremath{\psi^{#1}_{#2}}}

\newtheorem{theorem}{Theorem}

\newtheorem{corollary}[theorem]{Corollary}
\newtheorem{proposition}[theorem]{Proposition}

\newtheorem{lemma}[theorem]{Lemma}
\theoremstyle{remark}
\newtheorem{remark}[theorem]{Remark}
\makeatletter
\newcommand{\problemtitle}[1]{\gdef\@problemtitle{#1}}
\newcommand{\probleminput}[1]{\gdef\@probleminput{#1}}
\newcommand{\problemquestion}[1]{\gdef\@problemquestion{#1}}
\NewEnviron{problem}{
  \problemtitle{}\probleminput{}\problemquestion{}
  \BODY
  \par\addvspace{.5\baselineskip}
  \noindent
  \begin{tabularx}{\textwidth}{@{\hspace{\parindent}} l X c}
	  \multicolumn{2}{@{\hspace{\parindent}}l}{\textsc{\@problemtitle}} \\
    \textbf{Input:} & \@probleminput \\
    \textbf{Output:} & \@problemquestion
  \end{tabularx}
  \par\addvspace{.5\baselineskip}
}

\newtheorem{prealphthm}{{\bf Theorem}}

\newenvironment{alphthm}{\begin{prealphthm}\rm{ \hspace{-0.5  em}{\bf\ }}}{\end{prealphthm}}
	\newtheorem{predefin}[theorem]{{\bf Definition}}

\newenvironment{defin}[1]
{\begin{predefin}
		\rm{ \hspace{-0.5  em}{\bf\ }}{\rm #1}
		\hfill{$\blacktriangle$}}
	{\end{predefin}
}
\colorlet{color_new}{red!60!black} 
\colorlet{color_todo}{red!70!black} 

\def\isdef{\mbox {$\ \stackrel{\rm def}{=} \ $}}

\NewEnviron{new}{%
 \color{color_new} \ 
	\BODY 
	\color{black}
}
\newcommand{\ssc}{\scriptscriptstyle}

\def\DEBUG{defined}

\NewEnviron{todo}{%
	\ifdefined\DEBUG \color{color_todo} 
	\BODY \  
	\color{black}
	\fi
}

\title{Mean Isoperimetry with Control on Outliers: Exact and Approximation Algorithms}
\author{%
	Morteza Alimi  \\{\small Department of Mathematical Sciences},%
		\\ {\small Sharif University of Technology} \\ {\small morteza.alimi@student.sharif.edu } %
	\and Amir Daneshgar \thanks{corresponding author}\\{\small Department of Mathematical Sciences},%
		\\ {\small Sharif University of Technology} \\ {\small daneshgar@sharif.edu } %
	\and Mohammad-Hadi Foroughmand-Araabi \\{\small Department of Mathematical Sciences},%
		\\ {\small Sharif University of Technology} \\ {\small foroughmand@sharif.edu } }
\date{}

\begin{document}
\maketitle
\begin{abstract} 
	Given a weighted graph $G=(V,E)$ with weight functions $c:E\to \mathbb{R}_+$ and $\pi:V\to \mathbb{R}_+$, and  a subset $U\subseteq V$, the normalized cut value for $U$ is defined as 
	the sum of the weights of edges exiting $U$ divided by the weight of vertices in $U$.
	The {\it mean isoperimetry problem}, $\mathsf{ISO}^1(G,k)$, for a weighted graph $G$ is a generalization of the classical uniform sparsest cut problem 
	in which, given a parameter $k$,  the objective is to find $k$ disjoint nonempty subsets of $V$ minimizing the 
	average normalized cut value of the parts. The robust version of the problem seeks an optimizer where the number of vertices that fall out of the subpartition is bounded by 
	some given integer $0 \leq \rho \leq |V|$.
	
	Our main result states that $\mathsf{ISO}^1(G,k)$, as well as its robust version, $\mathsf{CRISO}^1(G,k,\rho)$, subjected to the condition that each part of the subpartition induces a connected subgraph,
	are solvable in time $O(k^2 \rho^2\ \pi(V(T)^3)$ on any weighted tree $T$, in which $\pi(V(T))$ is the sum of the vertex-weights. 
	This result implies that $\mathsf{ISO}^1(G,k)$ is strongly polynomial-time solvable on weighted trees when the vertex-weights are polynomially bounded and may be compared to the fact that  the problem is NP-Hard for weighted trees in general. 
	As far as applications are concerned, the connectivity requirement may be interpreted as an approach to model the practical consistency of the parts, which together with having control on the size of the outlier set and applying a smooth ``mean" cost function (as opposed to, say, the ``max" version), 
    characterizes our solution to $\mathsf{CRISO}^1(G,k,\rho)$ on trees as one of the most flexible and accurate procedures within the framework of isoperimetry-based clustering. 
	
	Also, using this, we show that both mentioned problems, $\mathsf{ISO}^1(G,k)$ and $\mathsf{CRISO}^1(G,k,\rho)$ as well as the ordinary robust mean isoperimetry problem $\mathsf{RISO}^1(G,k,\rho)$, admit polynomial-time $O(\log^{1.5}|V| \log\log |V|)$-approximation algorithms for weighted graphs 
	with polynomially bounded weights, using the R{\"a}cke-Shah tree cut sparsifier. To the best of our knowledge, this is the first polynomial-time  
	polylogarithmic approximation algorithm for $\mathsf{ISO}^1(G,k)$ in general (i.e.\  holding for all $k \geq  2$). 
	The result may be considered as a counterpart of a result of Arora, Lee and Naor (2008) indicating that the nonuniform sparsest cut problem, 
	as the other generalization of the uniform sparsest cut problem, admits a polynomial-time  $O(\sqrt{\log(|V|)} \log\log |V|)$-approximation algorithm 
	for all edge-weighted graphs (note that $\mathsf{ISO}^1(G,k)$ for $k=2$ is the uniform sparsest cut problem).
\end{abstract}
\thispagestyle{empty}

\section{Motivation and Background}\label{sec:introduction}

{\it Isoperimetry} is among the central topics in mathematical sciences, while the concept within the discrete setup is deeply related to graph theory, geometric optimization, and computer science 
(e.g.\  see \cite{CHA00,CHU97,DAJAMI12,daneshgar-javadi-pattern,GRI18,LIU15,naor2010l1,TER10} and references therein for general background). In particular,
different variants of the isoperimetry parameters may be interpreted as a class of invariants measuring {\it connectivity} or {\it density} of the whole space in a very general sense, while within the discrete setup, such parameters may be described as the optimum value of a graph partitioning or a clustering optimization problem, having close ties to such topics as {\it cut problems}, {\it flow problems} or {\it connectivity invariants} in graph theory, with a variety of applications in science and technology. Historically, the topic was introduced in its discrete setup in \cite{ALO86,ALMI85,JESI88,MIH89,MOH89} with motivations from communication theory seeking highly connected networks with some constraints (e.g.\  as regularity)
on the number of edges, giving rise to the theory of expander graphs in which proving the existence or constructing the extremal cases called Ramanujan graphs are still among deep and challenging problems in the field (e.g.\  see \cite{HOLIWI06,LUB94,MASPSR13-2,MASPSR15,MASPSR13} and references therein. Also, see \cite{daneshgar-javadi-jctb} for distinctions between the discrete and the continuous setups). On the other hand, and even in the discrete setting, the topic is closely related to the {\it concentration of measure} phenomenon and the theory of {\it metric embeddings}
with intimate connections to such fundamental conjectures and open problems as the {\it unique games conjecture}, {\it exponential time hypothesis} or theory of {\it zeta functions} (e.g.\  see \cite{IMPA01,IMPAZA01,MATR18,naor2010l1,TER10}).

To be more precise, let us consider a {\it weighted graph} $G = (V,E)$ with the
weight functions $c:E\to \mathbb{R}_+$ and $\pi:V\to \mathbb{R}_+$ on its sets of edges and vertices, respectively\footnote{Hereafter, the phrase ``weighted graph" refers to a graph with weights on both of its vertex and edge sets, in general. If one of these sets is not weighted, i.e.\  have trivial weights equal to one, then one may use the terms ``{\it edge-weighted graph}" or 
``{\it vertex-weighted graph}", accordingly. For weighted graphs with {\it potentials} see \cite{daneshgar-javadi-pattern}.}. 
Then, given any subset $A \subseteq V$, the {\it normalized cut} value corresponding to $A$ is defined as
$$\varphi(A) \isdef \frac{c(\delta(A))}{\pi(A)},$$
in which $\delta(A) \isdef \{e=uv \in E \ | \ u \in A \ \& \ v \not \in A\}$ is the set of boundary edges of $A$ and the weight of a set is the sum of the weights of its members. Let us also define $\mathcal{D}_{k,\rho}(V)$ to be the set of all $k$-tuples $\mathcal{A}=(A_{_{1}},A_{_{2}} \ldots, A_{_{k}})$ of nonempty disjoint subsets 
$A_{_{i}} \subseteq V$, called {\it $k$-subpartitions}, such that   
$$\displaystyle{\sum_{1 \leq i \leq k}} |A_{_{i}}| \geq |V|-\rho,$$
and, moreover, define $\mathcal{D}_{k}(V) \isdef \mathcal{D}_{k,|V|}(V)$ as well as the vector 
$$\Phi_{_{G,k}}(\mathcal{A}) \isdef [\varphi(A_{_{1}}), \varphi(A_{_{2}}), \ldots, \varphi(A_{_{k}})] \in \mathbb{R}^k,$$
along with the cost function,
$$ 	\Psi_{G,p}(\mathcal{A}) \isdef \|\Phi_{_{G,k}}(\mathcal{A})\|_{_{p}}.$$
Within this setup, one may define the following computational problems.
\begin{itemize}
	\item{{\bf The isoperimetry problem $\mathsf{ISO}^p(G,k)$}\\
	   {\bf Input}:  A weighted graph $G=(V,E,c,\pi)$ with integer weights and an integer $k > 1$.\\
	   {\bf Output}: $\mypsi{p}{k}(G) \isdef \displaystyle{\min_{\mathcal{A}\in \mathcal{D}_k(V)}} \Psi_{G,p}(\mathcal{A}).$
}
\item{{\bf The robust isoperimetry problem $\mathsf{RISO}^p(G,k,\rho)$}\\
	{\bf Input}:  A weighted graph $G=(V,E,c,\pi)$ with integer weights, along with integers $k > 1$ and $\rho \geq 0$.\\
	{\bf Output}: $\mypsi{p}{k,\rho}(G) \isdef \displaystyle{\min_{\mathcal{A}\in \mathcal{D}_{k,\rho}(V)}} \Psi_{G,p}(\mathcal{A}).$
}
\end{itemize}
\textit{Convention:} In the sequel, and without loss of generality, we always assume that the input graph $G=(V,E)$ is connected, 
and we usually let $n$ stand for the number of vertices $n$. Also, we restrict attention to mean and max versions of the isoperimetry problem, and 
whenever we use the superscript $p$ in an expression, we just mean that the expression is valid for both $p=1$ and $p=\infty$. 

	Our main objective in this paper is to study the computational complexity as well as the approximability of the problem $\mathsf{RISO}^1(G,k,\rho)$. Although the study of $\mathsf{ISO}^p(G,k)$,
	at least for $p \in \{1,\infty\}$  and $k=2$, is  a classic problem in spectral geometry of graphs  and  discrete geometric optimization, the problem for $k > 2$ has only mainly been studied in the past ten years or so, in relation to the study of higher Cheeger-type inequalities with motivations in geometric analysis, probability theory, graph theory, and computer science. Also, the robust version of the problem, $\mathsf{RISO}^p(G,k,\rho)$, has only been considered recently, and to the best of our knowledge,
\cite{Javadi-Ashkboos1} is the pioneering contribution starting a systematic study for the case $p=\infty$ 
	(also see \cite{daneshgar-javadi-pattern} and references therein for background about outlier detection).

In what follows, we first go through some necessary background and will try to provide a big picture showing the relative importance of our results before we proceed to the details in the forthcoming sections.

Since we are just considering discrete computational models (i.e.\  we are not working in any kind of {\it real} computational model), we always assume that the weights are rational 
numbers, and using finiteness of graphs along with multiplying by the greatest common divisor of the denominators, one may w.l.o.g.\  assume that the weights are positive integers.
	The isoperimetry problem $\mathsf{ISO}^p(G,k)$ is sometimes denoted by IPP, while the robust isoperimetry problem $\mathsf{RISO}^p(G,k,\rho=0)$, where one optimizes over the set of {\it partitions}, is a graph partitioning problem sometimes referred to as the {\it normalized cut problem}, denoted by NCP (e.g.\  see \cite{daneshgar-javadi-trees}). 
It is interesting to note that the set of vertices that are not contained in an optimal $k$-subpartition may be interpreted as a set of {\it outliers}, hence, providing an intuition 
	for $\mathsf{RISO}^p(G,k,\rho)$ as a problem related to clustering with a control on the number of outliers (e.g.\  see \cite{daneshgar-javadi-pattern,Javadi-Ashkboos1} and references therein for more on this).

Following the literature, 
	$\mathsf{ISO}^{\infty}(G,k)$ is the {\it classical} (maximum version) isoperimetry problem, where $\frac{1}{k}\mathsf{ISO}^{1}(G,k)$ is usually referred to as the {\it mean} isoperimetry problem\footnote{Note that our cost function is not the mean, but $k$ times what is usually called the mean isoperimetry cost function, e.g.\  as in \cite{daneshgar-javadi-trees}. Also, it ought to be mentioned that for $k=2$ the mean isoperimetry problem is sometimes referred to as the {\it uniform sparsest cut} or the {\it product sparsest cut} problem (e.g.\  see \cite{KRWA16}).}.

	Also, we adopt the notational standard used in \cite{daneshgar-javadi-trees}, and write constants as subscripts. For instance, $\mathsf{ISO}_{_{5}}^{\infty}(G)$ is the classical isoperimetry problem where the number of parts, $k=5$, is a constant and is not given as part of the input. Note that, if necessary, we may be more explicit and write $\mathsf{RISO}_{_{k=5,\rho=0}}^{\infty}(G)$
to distinguish between different constants of the problem. 
	For the sake of brevity, in the following we might occasionally drop the input part of a problem name, e.g.\ 
	we may write $\mathsf{RISO}^p$ instead of $\mathsf{RISO}^p(G,k,\rho)$, and likewise for $\mathsf{ISO}^{p}$. 

	Moreover, it is instructive to note that, $\mathsf{ISO}^p(G,k)$ is equivalent to $\mathsf{RISO}^p(G,k,n)$, while we also know that 
	the problems $\mathsf{RISO}_{_{k=2,\rho=0}}^{p}(G)$ and $\mathsf{ISO}_{_{2}}^{p}(G)$ are equivalent (see \cite{daneshgar-javadi-jctb} and Proposition~\ref{pro:basicineq}(c)), i.e.\  
	$$  \mathsf{RISO}_{_{k=2,\rho=0}}^{p}(G) \equiv {\rm \mathsf{ISO}}_{_{2}}^{p}(G),  $$
where such minimum costs are referred to as {\it sparsest cut} parameters, {\it edge expansion} or {\it Cheeger's constant} in the literature (e.g.\  see \cite{daneshgar-javadi-jctb,daneshgar-javadi-trees,DAJAMI12} and references therein).

Let $\mathcal{CD}_{k,\rho}(V) \subseteq \mathcal{D}_{k,\rho}(V)$ be the set consisting of $k$-subpartitions $\mathcal{A}=(A_{_{1}},A_{_{2}} \ldots, A_{_{k}})$ of $V$ having at most $\rho$ outliers such that for every 
	$i \in \{1,2,\ldots,k\}$ the induced subgraph $G[A_{_{i}}]$ is connected. Then we refer to the {\it connected} versions of the aforementioned isoperimetry problems by adding a capital C as a prefix to the corresponding acronym as in $\mathsf{CISO}^p(G,k)$ or $\mathsf{CRISO}^p(G,k,\rho)$. Also, we add a bar for the corresponding minimum costs as in $\bar{\psi}^{p}_{k}(G)$ or $\bar{\psi}^{p}_{k,\rho}(G)$.

	It is known that $\mathsf{ISO}^p(G,k)$ and $\mathsf{CISO}^p(G,k)$ are equivalent problems\footnote{Note that this is by no means true if we impose some bound on the number of outliers.} since $\mathsf{ISO}^p(G,k)$ always admits a connected minimizer \cite{daneshgar-javadi-trees}.
On the other hand, it is quite interesting to note that such a restrictive condition on the isoperimetry problem 
is, on the one hand, in concordance with our intuition looking for more consistent optimizers when using the problem in a clustering setup, 
	and on the other hand, it is known and will be shown again in this article that such a connectivity condition will lead to a decrease in computational complexity of the corresponding isoperimetry problem (also see \cite{Javadi-Ashkboos1} for such evidence).

Along the same lines of thought, one may consider the following variations of the isoperimetry setup and their effects on the computational complexity of the problem.
	\begin{itemize}
	\item[{\rm R1)}] Restriction to minimize over the set of $k$-partitions (i.e.\  $\rho=0$).
	\item[{\rm R2)}] Restriction to minimize over the set of $k$-subpartitions with an upper bound on the number of outliers.	
	\item[{\rm R3)}] Restriction to minimize over the set of connected $k$-subpartitions.
	\item[{\rm R4)}] Considering the classical case of $p=\infty$ instead of $p=1$.
\end{itemize}

Intuitively, it is folklore, that imposing (R1) or (R2), gives rise to problems at least as hard as the original one (usually harder),  
while  (R3) or (R4) usually gives rise to computationally easier problems. In this regard, it ought to be noted that a combination of restrictions (R2) and (R3) is the best choice for real applications which provides a very interesting trade-off that has been studied in \cite{Javadi-Ashkboos1} for the max version (i.e.\  $p=\infty$), and is studied in this article for the mean isoperimetry problem (i.e.\  when $p=1$). It is also instructive to mention that the performance of the 
	max isoperimetry cost-function in real applications may be quite poor due to the possibility of a large discrepancy between the maximum and the minimum of the normalized cut values of the parts in the problem's minimizers, justifying our study of the mean version from a practical point of view. Hence, among many other reasons discussed in what follows, this puts the problem $\mathsf{CRISO}^1(G,k,\rho)$ 
 at the top of the list to be considered for its practical and theoretical importance.

	Similarly, one may also consider the {\it decision} versions of such isoperimetry problems in which an extra rational number $\eta \in \mathbb{Q}_{+}$ 
	is assumed to be given in the input, and the query is to verify that the minimum cost function is less than or equal to $\eta$. In what follows a capital D as a prefix in the acronym of the problem indicates that we are referring to the decision version as, for instance, in $\mathsf{DCISO}^p(G,k,\eta)$ or in $\mathsf{DRISO}^p(G,k,\rho,\eta)$.

	Interpreting the normalized cut value $\varphi(A)$ as 
	$$\frac{c(\delta(A))}{w(A)}=\frac{\|\nabla\chi_{_{A}}\|_{_{1}}}{\|\chi_{_{A}}\|_{_{1}}}=\frac{\|\nabla\chi_{_{A}}\|^{^2}_{_{2}}}{\|\chi_{_{A}}\|^{^2}_{_{2}}}$$
	in which $\chi_{_{A}}$ is the characteristic function of the set $A$ and $\nabla$ is the gradient operator 
	(e.g.\  see \cite{daneshgar-javadi-jctb,DAJAMI12} for the definitions), has many interesting consequences%
	\footnote{Note that similar interpretations but using different norms in the numerator and the denominator are usually studied within the context of weighted Hardy inequalities as a study of lower-bounds for connectivity measures (or constants) (e.g.\  see \cite{EVPI95,MUC72,WAN94} for background).}.
	
	First, one may show that a Federer-Fleming-type theorem holds, indicating that the solution of the real-relaxed version of the isoperimetry problem for nonnegative real functions whose 	supports constitute a $k$-subpartition is the same as the solution of the discrete isoperimetry problem itself \cite{daneshgar-javadi-jctb,ROT85} (also see \cite{naor2010l1} for the case of the nonuniform sparsest cut problem). 
	
	On the other hand, an extreme real-relaxation of the $L_2$-norm version (i.e.\  letting function values also vary in sign) leads to the equality
	$$\frac{\|\nabla f\|^{^2}_{_{2}}}{\|f\|^{^2}_{_{2}}}=\frac{\langle \Delta f,f \rangle}{\langle f,f \rangle}$$
	providing an interpretation of the normalized cut value as some sort of normalized {\it energy} (i.e.\  Dirichlet) form. Moreover, clearly, within this geometric-analytic $L_2$-norm setup, minimizing over 
	all $k$-tuples of real functions, is essentially equivalent to computing the eigenvalues of the Laplacian operator $\Delta$  by the celebrated Courant-Fischer-Weyl variational formula. Hence, in this sense,
	the eigenvalues are the $L_2$-norm counterparts of the $L_1$-norm isoperimetry constants. 
	From this point of view, it is noticeable that the eigenvalue problem is actually polynomial-time solvable while the Cheeger-type\footnote{For background on Cheeger-type inequalities e.g.\  see \cite{CHE70,FUN13,MAZ60,MAZ61,MAZ62}.} inequalities may be interpreted as proofs for the approximation factor of the corresponding spectral algorithm for the isoperimetry problem
	(e.g.\  see \cite{CHU20,daneshgar-javadi-jctb,DAJAMI12,KWO13,LEOVTR14,LIU15,TRE13} and Table~\ref{tab:isofunction}).
	
	It is interesting that the spectral approach actually provides a geometrization of the isoperimetry problem in the sense that applying the Ng-Jordan-Weiss spectral procedure using the first $k$ eigenvectors of the Laplacian embeds the graph in $\mathbb{R}^k$, where one of the main parts of the proof provided in \cite{LEOVTR14} for higher 
	Cheeger inequalities justifies that the isoperimetry solution of the geometrized data embedded in $\mathbb{R}^k$ is a good approximation for the original isoperimetry problem on the given graph.
	

	A collection of what we know\footnote{This is to the best of our knowledge at the time of writing and limited to the topics discussed in this article.} 
	about the computational complexity of all variants of the isoperimetry problem is summarized in Table~\ref{tab:isofunction}. It must be noted that clustering, in general, is a hard problem mainly for the existence of instances that are not well-separated (for more on this and a study of phase-transitions in the stochastic block model see e.g.\  \cite{ABB18} and references therein). Hence, our general expectation is to face a strong computational hardness for different versions of the isoperimetry problem at least when one is considering the problem in a general setting. Adding in the crucial importance of the clustering problem both from theoretical and practical points of view, this computational hardness makes the whole study more interesting and also quite challenging. In this regard, seeking approximation algorithms is quite justified from both theoretical and practical points of view, as we shall see in Section~\ref{sec:general} that although the behavior of these different variants turn out to be quite different when one tries to find exact values of the minimum costs, they are essentially computationally equivalent  up to  constant factor approximation algorithms.
	
	There are some rare cases when one may efficiently solve a version of the isoperimetry problem. Not only such cases may crop up as bits of surprise, but also such instances of the problem usually contain 
	some new aspects or techniques to be applied in further applications and studies. Among such efficiently solvable cases, one may refer to the solvability of $\mathsf{DISO}^{\infty}(G,k,\eta)$
	on weighted trees in linear time\footnote{Note that in a {\it weighted graph} model both vertices and edges are weighted, in general.}. There are two aspects to this bit of a surprise that ought to be mentioned.
	
	Considering different parameters of an isoperimetry problem, one may note that pushing the problem toward the maximum norm as well as making the graph as sparse as a tree are the most important means to make the problem efficiently solvable; however, having arbitrary weights on both vertices and edges still makes the whole solution nontrivial. Here comes the main surprise which is deeply related to an extension of the whole theory to the case of weighted graphs with potentials, or in other words,
	extending the whole theory from the setup of ordinary Laplace operators to Schr\"odinger operators on graphs (e.g.\  see \cite{BECAEN05,daneshgar-javadi-pattern,raminthesis}). Specifically,  although an easy and efficient algorithm for the problem is not obvious at all in the ordinary setup, the efficient solution to the problem easily crops up as a bottom-up traversal of the tree itself in which 
	the whole data is transferred upwards through the potential function. This strong evidence, in particular, justifies the importance of a thorough study of isoperimetry problems within the setup of graphs with potentials (see also Remark~\ref{rem:crisomax}). 
	
	On the other hand, having a variety of parameters in an isoperimetry problem with a direct impact on the computational complexity of the problem at hand provides an opportunity 
	to construct and study variants of this problem that are quite close and on opposite sides of the P~{\it vs.}~NP borderline. To see this, note that,  $\mathsf{DISO}^{1}(G,k,\eta)$ is NP-complete even on weighted trees (also compare to the maximum version which is linear-time solvable on weighted trees \cite{daneshgar-javadi-trees}), while one may try to push the problem a bit further by optimizing only over {\it connected subpartitions} and investigate the computational complexity of this more restricted version. 
	
	As a summary of our main results, we may mention that this extra bit of restriction not only turns $\mathsf{CRISO}^{1}(G,k,\rho)$ into a polynomial-time solvable problem on weighted trees with polynomially bounded vertex-weights, but, surprisingly, it also provides a back-door to provide polynomial-time approximation algorithms for all variants of the isoperimetry problem discussed here on a general weighted graph with polynomially bounded weights (see Section~\ref{sec:mainresults} for more details).
	

	Considering the fact that clustering and partitioning are among the most fundamental problems in modern science and technology, 
	it is quite natural to see a large amount of scientific research focused on the approximability of such problems, in view of their computational hardness. 
	Although within the theoretical realm the study was mainly initiated by the classical (i.e.\  maximum version) sparsest cut problem $\mathsf{ISO}_{_{2}}^{\infty}(G)$ (e.g.\  see \cite{ALMI85,JESI88,MIH89,MOH89} for background), 
	among the best pioneering contributions in the approximation-algorithmic side, there is a result of Arora, Rao, and Vazirani
	\cite{ARRAVA09} that provides an $O(\sqrt{\log n})$-approximation algorithm for $\mathsf{ISO}_{_{2}}^{1}(G)$ based on an SDP relaxation (also see \cite{AURA98,LERA99,LILORA95} for background and \cite{ARLENA08,DEKHSAVI06,naor2010l1} for more on this and the general nonuniform sparsest cut problem). 
	
	On the other hand, as mentioned before,  a classical or higher-order Cheeger-type inequality that bounds the isoperimetric constant by eigenvalues of the Laplacian as
	$$O(\lambda_{_{k}}) \leq \mypsi{p}{k} \leq O(f(k)\sqrt{\lambda_{_{k}}}),$$
    for some function $f$, may be interpreted as a proof of the fact that a polynomial-time spectral algorithm computing $\lambda_{_{k}}$ is a $O(1/\sqrt{\lambda_{_{k}}})$-approximation algorithm for 
    $\mypsi{p}{k}$. This interpretation initiated in \cite{daneshgar-javadi-jctb,daneshgar-javadi-trees,raminthesis,OVTR11} (also see \cite{CHRetal11}), is called a {\it Cheeger-type approximation} by Trevisan (see \cite{TRE13} and references therein), referring to any approximation algorithm for the isoperimetry problem providing  an $O(\sqrt{OPT})$ solution.
    This approach eventually gave rise to a proof of a higher Cheeger inequality in \cite{LEOVTR14} for $\mypsi{\infty}{k}$ as (also see \cite{DAJAMI12} for the case of weighted cycles where $f(k)$ is a constant and \cite{KWO13} for inequalities of more general forms),
	$$O(\lambda_{_{k}}) \leq \mypsi{\infty}{k} \leq O(k^2\sqrt{\lambda_{_{k}}}).$$

	Our second main result in this article is the fact that essentially all variants of the isoperimetry problem discussed before are equivalent as far as approximation is concerned (disregarding constant factors, say, less than $8$),
	while we use a tree cut-sparsifier of R{\"a}cke and Shah \cite{racke2014improved} to show that all these problems are polynomial-time 
	$O(\log^{1.5} n \log\log n)$-approximable on general weighted graphs with polynomially bounded weights (see Section~\ref{sec:general} for more details).
	
	 In this regard, and for the sake of juxtaposition, it is quite interesting to have a look at the nonuniform sparsest cut problem as a sister to the isoperimetry problem (e.g.\  see \cite{ARLENA08,DEKHSAVI06,naor2010l1} and references therein). It is 
	 well-known that the nonuniform sparsest cut for equal demands (often called the {\it uniform sparsest cut} problem (e.g.\  see \cite{KRWA16}) is essentially the same problem as $\mathsf{ISO}_{_{2}}^{1}(G)$, while a celebrated result of Arora, Lee, and Naor \cite{ARLENA08}, based on metric embedding techniques, indicates that the nonuniform sparsest cut problem admits  a polynomial-time  $O(\sqrt{\log n} \log\log n)$-approximation on edge-weighted graphs\footnote{It must be pointed out that the setup of such a problem is basically the setup of edge-weighted graphs while the role of demands may be compared to the role of vertex-weights in isoperimetry. Hence, the uniform sparsest cut problem is essentially $\mathsf{ISO}_{_{2}}^{1}(G)$ on edge-weighted graphs, which is the setup of approximation results in \cite{ARLENA08,ARRAVA09}. This is quite a subtle observation when one notes, e.g.\  in relation to a main result of this article, that $\mathsf{ISO}_{_{2}}^{1}(G)$ is NP-complete on weighted trees but is polynomial-time solvable if the vertex-weights of the tree are polynomially bounded.}.
	 On the other hand, although we know a lot about the hardness or inapproximability of the general nonuniform sparsest cut problem (e.g.\  see \cite{CHW06,GUTAWI13,KHVI15,naor2010l1,TRE13}), it is quite intriguing that we do not even know whether a subexponential-time $O(1)$-approximation algorithm  exists for the uniform case $\mathsf{ISO}_{_{2}}^{1}(G)$ (e.g.\  see \cite{MATR18,naor2010l1} for more on this\footnote{It should be noted that there are a number of results in the literature concerning approximating these problems subject to a variety of {\it well-separatedness} conditions such as eigenvalue bounds or changing the number of parts as in \cite{louis2014approximation}, which are not relevant to the subject of our study in this article.}).


\begin{table} \scriptsize
	\caption{\protect\label{tab:isofunction}Summary of known results about variants of the isoperimetry problem}
	\centering
	\begin{tabular}{ |l|l|l|l| }
	\hline \hline
	\textbf{Problem} & \textbf{Restrictions} & \textbf{Time Complexity} & \textbf{Ref./Comments} \\ \hline \hline
	
		\multirow{2}{*}{$\mathsf{ISO}_{_{k}}^{p}(G)$} & - weighted graphs  & NP-Hard & \cite{daneshgar-javadi-trees}\\
        &- weighted bipartite with & NP-Hard & \cite{daneshgar-javadi-trees}\\
        & \quad tree-width $\leq 2$ &  & \\
      	&- unweighted graphs & NP-Hard &  \cite{daneshgar-javadi-trees}\\
      	&- weighted trees & polynomial-time & \cite{daneshgar-javadi-trees}\\  
		\multirow{2}{*}{$\mathsf{ISO}_{_{k}}^{1}(G)$} &- weighted trees & polynomial-time & \cite{daneshgar-javadi-trees}\\
        &  & $(k+\epsilon)$-approx. & for $k=2$ also see \cite{MASH90}\\  
   		&- edge-weighted graphs & polynomial-time & \cite{ARRAVA09} (also see \cite{LERA99,WCMS04})\\   
        &  & $O(\sqrt{\log n})$-approx. &  \\                                                    	   
		$\mathsf{ISO}_{_{k}}^{\infty}(G)$&- weighted trees & FPTAS &  \cite{daneshgar-javadi-trees}\\   
         	&  &  & \\
	\hline
	\hline		
		\multirow{2}{*}{$\mathsf{ISO}^{1}(G,k)$} &- weighted trees & NP-Hard & \cite{daneshgar-javadi-trees}\\	
        &- weighted trees & strong polynomial-time & this article: Theorem~\ref{thm:iso}\\
		&\ \  with polynomially   &   & same solution \\
		&\ \ bounded vertex-weights  &  & exists for $\mathsf{CISO}^1(G,k,\rho)$\\			       
		&- weighted graphs  & polynomial-time  & this article: Theorem~\ref{thm:mainapprox} \\
		&\ \  with polynomially   &  $O(\log^{1.5}n\log\log n)$-approx.  & same solution\\
		&\ \ bounded weights  &  & exists for $\mathsf{RISO}^1(G,k,\rho)$\\			
		$\mathsf{DISO}^{\infty}(G,k,\eta)$ &- weighted trees & $O(n)$ & \cite{daneshgar-javadi-trees} \\
				&  &  & (see \cite{daneshgar-javadi-pattern} for $\mathsf{ISO}^{\infty}(G,k)$) \\	      
              	&  &  & (see \cite{MOH89} for $k=2$)\\	       
	\hline		
	\hline		
		\multirow{2}{*}{$\mathsf{RISO}_{_{k=2,\rho=0}}^{\infty}(G) \equiv {\rm \mathsf{ISO}}_{_{k=2}}^{\infty}(G)$} &- unweighted graphs  & NP-Hard & \cite{MOH89}\\
                        	&\ \  with multiple edges &  & \\
		\multirow{1}{*}{$\mathsf{RISO}_{_{k=2,\rho=0}}^{1}(G) \equiv {\rm \mathsf{ISO}}_{_{k=2}}^{1}(G)$} &- weighted graphs  & NP-Hard & \cite{MASH90}\\   
        &- bipartite planar  & NP-Hard & attributed to \\
		&\ \ weighted graphs &  & Papadimitrou 1997 in \cite{SHMA00}  \\
		\multirow{1}{*}{$\mathsf{RISO}_{_{k,\rho=0}}^{p}(G)$} & - weighted graphs  & NP-Hard & \cite{daneshgar-javadi-trees}\\
        &- weighted bipartite with & NP-Hard &  \cite{daneshgar-javadi-trees}\\
        & \quad tree-width $\leq 2$ &  & \\
        &- unweighted graphs & NP-Hard &  \cite{daneshgar-javadi-trees}\\	
		\multirow{1}{*}{$\mathsf{RISO}_{_{k,\rho=0}}^{\infty}(G)$} &- weighted trees & polynomial-time exact \& & \cite{daneshgar-javadi-trees}\\
        &  &$(k-1+\epsilon)$-approx.  & \\
		$\mathsf{RISO}_{_{k=3,\rho=0}}^{\infty}(G)$   &- weighted trees & $O(n^2)$ & \cite{daneshgar-javadi-trees}\\		
		\multirow{2}{*}{$\mathsf{RISO}_{_{k,\rho=0}}^{1}(G)$} &- weighted trees & polynomial-time exact \& & \cite{daneshgar-javadi-trees}\\
        &  & $(2k-2+\epsilon)$-approx. & \\       
		\multirow{1}{*}{$\mathsf{RISO}_{_{\rho=0}}^{\infty}(G,k)$} &- weighted trees & strongly NP-Hard & \cite{daneshgar-javadi-trees}\\
        &- unweighted trees & NP-Hard & \cite{daneshgar-javadi-trees}\\   
 	       	&  &  & \\             
		\multirow{1}{*}{$\mathsf{RISO}_{_{\rho=0}}^{1}(G,k)$} &- weighted trees & NP-Hard & \cite{daneshgar-javadi-trees}\\
 	       	&  &  & \\
		\multirow{1}{*}{$\mathsf{RISO}^{1}(G,k,\rho)$}&- weighted trees  &polynomial-time  2-approx.  & this article: Corollary~\ref{cor:2alpha}\\
        &\ \  with polynomially   &   & \\
        &\ \ bounded vertex-weights  &  & \\			            
        &- weighted graphs  & polynomial-time  & \\
        &\ \  with polynomially   &  $O(\log^{1.5}n\log\log n)$-approx.  & this article: Theorem~\ref{thm:mainapprox}\\
        &\ \ bounded weights  &  & \\			       
	\hline	
	\hline		
    	&  &  & \\
	\multirow{2}{*}{$\mathsf{CRISO}^{1}(G,k,\rho)$} &- weighted trees & $O(\rho^2k^2n^3 W^3)$ & this article: Theorem~\ref{thm:iso}\\
     &\ \  with polynomially bounded  &  &  $W=$ maximum vertex-weight\\
     &\ \  vertex-weights  &  &\\		
     &- weighted graphs  & polynomial-time  & \\
     &\ \  with polynomially   &  $O(\log^{1.5}n\log\log n)$-approx.  &  this article: Theorem~\ref{thm:mainapprox}\\
     &\ \ bounded weights  &  & \\			        	
	\multirow{1}{*}{$\mathsf{DCRISO}^{\infty}(G,k,\rho,\eta)$} &- weighted trees & $O((\rho+1)^2k^2n)$ & \cite{Javadi-Ashkboos1}\\
               &  &  & \\
	\hline

\end{tabular}	
\end{table}

\section{Main results}\label{sec:mainresults}

We have two main results in this paper regarding the computational complexity of the mean isoperimetry problem. Our first main result is a proof of the polynomial-time computability 
of $\mathsf{ISO}^{1}(T,k)$ where $T=(V,E)$ is a weighted tree whose vertex-weight function $\pi:V\to \mathbb{Q}_+$ is polynomially bounded i.e.\  
$$\forall\ u \in V, \quad \pi(u) \leq {\sf p}(n),$$
in which ${\sf p}(x)$ is a polynomial function of $x$ (see Theorem~\ref{thm:iso}). 
Note that this fact is quite nontrivial since we already know that $\mathsf{ISO}^{1}(T,k)$ is NP-complete on weighted trees in general \cite{daneshgar-javadi-trees}. 
The result essentially settles the computational complexity of the mean isoperimetry problem on weighted trees as one of the most interesting problems 
dancing on the edge of the border between the classes P and NP. 

Our main algorithm for solving $\mathsf{CRISO}^{1}(G,k,\rho)$ in polynomial time on weighted trees with polynomially bounded vertex-weights is, 
of course, comparable to the main result of \cite{Javadi-Ashkboos1} 
(also see \cite{daneshgar-javadi-trees,daneshgar-javadi-pattern}) proving that $\mathsf{CRISO}^{\infty}(G,k,\rho)$ is (weakly) polynomial-time solvable on weighted trees, showing a trade-off 
between going from max to mean version to increase the hardness, on the one hand, and relaxing the model by imposing the condition of polynomial-boundedness of the vertex-weights to decrease hardness, on the other hand\footnote{Actually, our algorithm's runtime in terms of the vertex-weights is pseudopolynomial, suggesting that the problem is not strongly NP-Hard.}.
To be more precise, let us mention that our algorithm's worst-case run-time is $O(k^2 \rho^2 n^3 W^3)$, where $W$ is the maximum of the vertex-weights. 
Independence of the running time from the edge-weights shows that our algorithm is strongly polynomial with respect to the edge-weights while it is pseudopolynomial with respect to the vertex-weights. 

Crucial to our approach are two manipulations, one on the problem itself by focusing on $\mathsf{CISO}^{1}(T,k,\rho)$ instead of $\mathsf{ISO}^{1}(T,k,\rho)$, and the other 
by manipulating the cost function in a way that it fits into a dynamic programming recursive procedure. 

It is instructive to note that, by what was already discussed in Section~\ref{sec:introduction},
one may expect that the problem $\mathsf{CISO}^{1}(T,k,\rho)$ be easier to solve than the original mean isoperimetry problem, while by the following result of \cite{daneshgar-javadi-trees} we already know that the answer to both problems is the same.
\begin{alphthm}[\cite{daneshgar-javadi-trees} Theorem 1] 
	\label{thm:daneshgarjavadi-connected}
	For any given weighted graph $G$ we have, $\bar{\psi}_{k}^p(G) = \psi_{k}^p(G)$.
\end{alphthm} 

On the other hand, we will see that in order to apply a recursive computational procedure, one has to somehow generalize the cost function in a way to fit it into the recursive structure of a tree as well as to make it possible to have good control on different possible cases appearing in the recursive procedure itself (see Definitions~\ref{def:modify1}, \ref{def:modify2}, \ref{def:modify3}   and the paragraph preceding Definition~\ref{def:modify2}). This approach, when properly applied, leads to a complete characterization of the optimal value as will be presented in Theorem~\ref{thm:tree}.

Another aspect of our first main result (see Theorem~\ref{thm:iso}) is that solving the robust version of the mean isoperimetry problem, 
$\mathsf{CRISO}^{1}(T,k,\rho)$, is in principle a much stronger result than solving the unrestricted mean isoperimetry problem itself%
\footnote{Note that Theorem~\ref{thm:daneshgarjavadi-connected} is not necessarily true for the robust cost function.}. 
Hence, from a practical point of view, our result provides a feasible solution to the mean isoperimetry problem in the best case possible as far as 
practical issues are concerned, by considering the smoothing property of the mean cost function, giving rise to much better solutions, 
and providing the best possible control on the size of the outliers in the dataset%
\footnote{We are not going to discuss practical issues in this article, however, one may also consider the trade-off existing between the quality of the solutions, 
runtime, and the approximation ratio of the existing algorithms. Also, note that for practical purposes one may start with very fast algorithms like the one 
existing for $\mathsf{ISO}^{\infty}(T,k)$ and gradually reduce the dataset to smaller configurations on which more precise procedures as our algorithm may be applied. For more on relations between the optimizers of an isoperimetry problem on a weighted graph and those of a minimum spanning tree of the graph itself e.g. see \cite{daneshgar-javadi-pattern} and references therein.}.

Our second piece of contribution in this article is to prove that all discussed varieties of isoperimetry problems admit polynomial-time $O(\log^{1.5} n \log\log n)$-approximation
algorithms on weighted graphs of polynomially bounded weights (see Theorem~\ref{thm:mainapprox}).  
In this regard, such approximation algorithms for the ordinary isoperimetry problems follow 
from known inequalities provided in \cite{daneshgar-javadi-trees} while we show that similar inequalities also hold for the connected versions of the problem (see Proposition~\ref{pro:connected}).
Our basic result, which constitutes the core of all these approximation algorithms, is an approximation algorithm for $\mathsf{ISO}^{1}(G,k)$ 
which is based on an application of our previously discussed exact algorithm on a R{\"a}cke-Shah's tree cut sparsifier. 


\section{Mean isoperimetry on weighted trees}
\label{sec:trees}

In this section, our main objective is to show that $\mathsf{CRISO}^{1}(G,k,\rho)$ is solvable on weighted trees in pseudopolynomial-time, 
and moreover, we will show that when the vertex-weights are polynomially bounded, our algorithm runs in strongly polynomial time (i.e.\ its running time does not depend on edge weights).

In the sequel, we always assume that $T$ is an ordered rooted tree, i.e.\ an arbitrary vertex, say  $r_{_{0}}\in V(T)$, is specified as a root, and there is an ordering
over the children of any vertex of $T$. Also, for $r\in V(T)$, we let $T_r$ stand for the subtree of $T$ hanging from $r$. Moreover, for a vertex $r$ of $T$, 
the number of $r$'s children is denoted by $\textsf{lastChild}(r)$
and we typically denote the ordered list of the children of $r$ by $(v_1, v_2, \ldots, v_\textsf{lastChild}(r))$.
In this setting, the parent of $r$ is usually denoted by $\hat{r}$ and $T_{r,i}$ is used to denote the graph induced on the vertex set consisting of the vertex $r$ along with the vertex sets of $T_{v_1}, \ldots, T_{v_i}$, where $i\le \textsf{lastChild}(r)$. 
In other words, $T_{r,i}$ is the tree formed from $T_r$ by cutting off the trees $T_{v_{i+1}}, \ldots, T_{v_\textsf{lastChild}(r)}$. 

As mentioned before, in order to be able to derive a recursive formula for $\psi_{k,\rho}^1(T_r)$'s, we need to modify the cost function in two stages that will 
be explained in our following definitions and facts.

Firstly, let us define a more refined family of subpartitions of the vertex set of a tree.

\begin{figure}
	\label{fig:T_r}
	\begin{tikzpicture}[y=-1cm]
	\filldraw[black] (13.49778,10.54889) ellipse (0.09778cm and 0.1cm);
	\filldraw[black] (16.04222,10.54889) ellipse (0.09778cm and 0.1cm);
	\filldraw[black] (10.21333,10.54889) ellipse (0.09778cm and 0.1cm);
	\filldraw[black] (12.67556,7.84) ellipse (0.09778cm and 0.1cm);
	\filldraw[black] (11.60889,10.54889) ellipse (0.09778cm and 0.1cm);
	\draw[black] (12.67556,7.84) -- (10.21333,10.54889);
	\draw[black] (12.67556,7.84) -- (16.04222,10.54889);
	\definecolor{penColor}{gray}{0.8}
	\draw[penColor] (10.21333,10.30222) -- (9.47333,11.86222) -- (10.78667,11.86222) -- (10.21333,10.38444);
	\draw[penColor] (11.60889,10.30222) -- (10.86889,11.86222) -- (12.18222,11.86222) -- (11.60889,10.38444);
	\draw[penColor] (13.49778,10.30222) -- (12.75778,11.86222) -- (14.07111,11.86222) -- (13.49778,10.38444);
	\draw[penColor] (16.04222,10.30222) -- (15.30222,11.86222) -- (16.61556,11.86222) -- (16.04222,10.38444);
	\draw[black] (12.67556,7.84) -- (11.60889,10.54889);
	\draw[black] (12.67556,7.84) -- (13.49778,10.54889);
	\draw[penColor] (12.67556,7.59333) -- (14.64444,12.43778) -- (8.32444,12.43778) -- cycle;
	\draw[penColor] (12.51111,7.1) -- (6.6,12.92889) -- (19.57111,12.92889) -- cycle;
	\path (9.96667,11.04) node[text=black,anchor=base west] {$v_1$};
	\path (11.34444,11.04) node[text=black,anchor=base west] {$v_2$};
	\path (13.25111,11.04) node[text=black,anchor=base west] {$v_i$};
	\path (15.79556,11.04) node[text=black,anchor=base west] {$v_{\text{lastChild}(r)}$};
	\path (12.18222,10.62889) node[text=black,anchor=base west] {$\cdots$};
	\path (14.31778,10.71333) node[text=black,anchor=base west] {$\cdots$};
	\path (12.471111,8.41333) node[text=black,anchor=base west] {$r$};
	\definecolor{textColor}{gray}{0.8}
	\path (13.16889,12.82889) node[text=textColor,anchor=base west] {$T_r$};
	\path (11.28,12.35778) node[text=textColor,anchor=base west] {$T_{r,i}$};
	\end{tikzpicture}%
	\caption{Tree terminology illustration.}
\end{figure}

\begin{defin}{\label{def:modify1}
		Let $\mathcal{A} = \{ A_1, \ldots, A_k \} $ be a subpartition of $T_{r,i}$. We denote the weight of the part containing $r$ by $\omega(\mathcal{A})$; 
		in other words, 
		\begin{align*}
			\omega(\mathcal{A}) \isdef &
			\begin{cases}	
				\pi(A_j) & {\rm if\ } r\in A_j\ {\rm for\ some\ } 1 \leq j \leq k, \\
				0 & \not \exists 1 \leq i \leq k, \quad r \in A_i.
			\end{cases}
		\end{align*}
Also, for fixed parameters $1 < k$, $0\le \omega \le \pi(V(T_{r,i}))$, and $\mathsf{b}\in \{ \textsf{true}, \textsf{false} \}$,
	define  $\mathcal{CD}_{k,\omega,\rho,\mathsf{b}}(T_{r,i})$ to be the family of all $k$-subpartitions $\mathcal{A} = \{A_1, \ldots, A_k\}$ of $V(T_{r,i})$, 
	satisfying the following conditions,
	\begin{enumerate}
		\item 
		The graph induced on each $A_i$ is connected for $1 \leq i \leq k$.
		\item
		$\omega(\mathcal{A})=\omega$.
		\item
		$ \mathsf{b}=\textsf{true} \ \ \Leftrightarrow \ \ \exists\  1 \leq j \leq k, \quad  \{r,v_i\} \subseteq A_j. $
		\item
		$|T| - \sum_{i=1}^{k} |A_i| \le \rho$. 
		
	\end{enumerate}
}\end{defin}

Our second modification concerns the recursive procedure on the tree itself. To see the issue, suppose that
 $\psi^1_k(T_{r,i})$ is minimized by the subpartition $\mathcal{A}$, and $r\in A_l \in \mathcal{A}$. As mentioned before, we are only considering connected parts,
and consequently, one may assume that each $A_j$ induces a connected subgraph. Therefore, with the possible exception of $A_l$, each $A_j$ lies completely either in 
$T_{r,i-1}$ or $T_{v_i}$. 
Let $A'_l \isdef A_l\cap T_{r,i-1}$ and $A''_l \isdef A_l\cap T_{v_i}$,  and also define,
$$\mathcal{A}' \isdef \{A'_l\} \cup \{A\in \mathcal{A} \mid A\subset T_{r,i-1} \},$$
and
$$\mathcal{A}'' \isdef \{A''_l\} \cup \{A\in \mathcal{A} \mid A\subset T_{v_i} \}.$$
Then if it were the case that $\mathcal{A}'$ and $\mathcal{A}''$ were optimum solutions for 
$\psi^1_{k,\rho}(T_{r,i-1})$ and $\psi^1_{k,\rho}(T_{v_i})$, respectively,
we could find (or algorithmically build) $\mathcal{A}$ inductively (or recursively) using $\mathcal{A}'$ and $\mathcal{A}''$. 
But $A_l$ (the part containing the root) foils this approach, since the contribution of $A_l$ to the objective function
(which is $\frac{c(\delta(A_l))}{\pi(A_l)}$) does not equal the sum of the contributions of $A_l'$ and $A_l''$. 

To overcome this issue, we introduce a new  parameter $\mathsf{d}$, and modify the denominator of the term corresponding to the root's normalized cut expression as follows.

\begin{defin}{{\bf Modified mean isoperimetry cost function}\label{def:modify2}
	
	Let $T$ be a rooted tree with root $r$ and 
	let $\mathsf{d}$ be an arbitrary integer. Define 
	
	\[
	\begin{array}{rcl}
		\Psi^{\mathsf{d}}_T(\mathcal{A}) &:=&  \frac{c(\delta(A_i))}{\mathsf{d}} + \sum_{j\ne i} \frac{c(\delta(A_j))}{\pi(A_j)} \\
		&=& \frac{c(\delta(A_i))}{\mathsf{d}} + \Psi_T(\mathcal{A}\setminus \{ A_i \}),
	\end{array}
	\]
	
	where $A_i$ is the part containing $r$ (if any). We might drop the $T$ subscript when the tree is clear from the context.   
}\end{defin}



Our first observation is the fact that Definitions~\ref{def:modify1} and \ref{def:modify2} provide a well-defined modified version of the cost function.
\begin{lemma}
	\label{lem:psiPsi}
	$$\bar{\psi}_{k,\rho}^1(T_{r,i}) = \min_{\omega,\mathsf{b}} \ \min_{\mathcal{A}\in\mathcal{CD}_{k,\omega,\rho,\mathsf{b}}(T_{r,i})}\Psi^{\omega}_{T_{r,i}}(\mathcal{A}),$$
	where the first minimum is taken over all 
	$0\le \omega \le \pi(V(T_{r,i})), \ 0\le \rho' \le \rho$ and $\mathsf{b}\in \{\textsf{true}, \textsf{false}\}$.
\end{lemma}
\begin{proof}
	For each $\mathcal{A}\in\mathcal{CD}_{k,\omega,\rho,\mathsf{b}}(T_{r,i})$, we have 
	$$\Psi^{\omega}_{T_{r,i}}(\mathcal{A}) = \Psi_{T_{r,i}}(\mathcal{A}).$$
	
	Also, 
	$$ \mathcal{CD}_{k,\rho}(V(T_{r,i})) =  \bigcup_{\substack{\ssc 0\le \omega\le \pi(V(T_{r,i})), \\  \ssc \mathsf{b}\in \{\textsf{true},\textsf{false}\}}} \mathcal{CD}_{k,\omega,\rho,\mathsf{b}}(T_{r,i}). $$
	
	Hence 
	\[
	\bar{\psi}_{k,\rho}^1(T_{r,i}) = \min_{\mathcal{A}\in \mathcal{CD}_{k,\rho}(V(T_{r,i}))} \Psi_{T_{r,i}}(\mathcal{A})
	= \min_{\omega,\mathsf{b}} \ \min_{\mathcal{A}\in\mathcal{CD}_{k,\omega,\rho,\mathsf{b}}(T_{r,i})}\Psi^{\omega}_{T_{r,i}}(\mathcal{A}),
	\]
	with the first minimum being taken over all 
	$0\le \omega \le \pi(V(T_{r,i})), \  \mathsf{b}\in \{\textsf{true}, \textsf{false}\}$.
	
\end{proof}

Motivated by the inner minimization expression in Lemma~\ref{lem:psiPsi}, we are now ready to define the generalized isoperimetry parameter, 
which forms our algorithm's primary object of computation. 

\begin{defin}{\textbf{Generalized mean isoperimetry parameter}
	\label{def:modify3}

	The generalized mean isoperimetry parameter is defined as,
		$$\bar{\psi}_{k,\omega,\mathsf{d},\rho,\mathsf{b}}(T_{r,i}) \isdef 
		\min_{\mathcal{A}\in\mathcal{CD}_{k,\omega,\rho,\mathsf{b}}(T_{r,i})}\Psi^{\mathsf{d}}_{T_{r,i}}(\mathcal{A}).$$

}\end{defin}

To put it more verbosely, $\bar{\psi}_{k,\omega,\mathsf{d},\rho,\mathsf{b}}(T_{r,i})$ denotes the minimum modified mean isoperimetry cost of $T_{r,i}$ with parameter $\mathsf{d}$
where the minimization is taken over all $k$-subpartitions of $T_{r,i}$ which induce connected parts, where there are at most $\rho$ outliers, 
the part containing $r$ has size $\omega$, and $r$ and $v_i$ are in the same part iff $\mathsf{b} = \textsf{true}$. 
Note that if $\omega=0$, then the values of $\mathsf{d}$ and $\mathsf{b}$ are immaterial in $\bar{\psi}_{k,\omega,\mathsf{d},\rho,\mathsf{b}}(T_{r,i})$; so in the following we assume
they are 0 and $\textsf{false}$, respectively, in this case. 

The following proposition 
shows the relationship between the generalized and standard versions of the (robust, connected) isoperimetry function which is an immediate consequence of definitions and Lemma~\ref{lem:psiPsi}. 

\begin{proposition}
	\label{prop:psilongpsi}
	\[ 
	\bar{\psi}^1_{k,\rho}(T_{r,i}) = \min\{ \bar{\psi}_{k,\omega,\omega,\rho,\mathsf{b}}(T_{r,i}) \mid 0\le \omega \le \pi(V(T_{r,i})), %
	\mathsf{b}\in \{\mathsf{true}, \mathsf{false}	\} \}.
	\]
\end{proposition}

Now we set out to prove our main result i.e.\  prove a recurrence relation for  $\bar{\psi}_{k,\omega,\mathsf{d},\rho,\mathsf{b}}$, that readily gives rise to 
a dynamic programming algorithm.


\textit{Convention}. In the following, when we are computing on a  subtree $T_v$ of a rooted tree $T$, to simplify cases and the presentation, we take the convention that the edge connecting $v$ to its parent in $T$ is also counted as an outgoing edge from $v$ (i.e.\ ``half'' of the edge $v\hat{v}$ belongs to $T_v$). 

\begin{theorem}
	\label{thm:tree}
	Let $T$ be a rooted tree with an ordering on the children of any node. 
	Moreover, let $r\in V$, $1\le i\le \mathsf{lastChild}(r)$, $1\le k\le |V(T_{r,i})|$,  $0\le \omega,\mathsf{d} \le \pi(V(T_{r,i}))$,  $0\le \rho \le |V(T_{r,i})|$, and $\mathsf{b}\in \{ \mathsf{true}, \mathsf{false} \}$ be given parameters.  Then 
	$\bar{\psi}_{k,\omega,\mathsf{d},\rho,\mathsf{b}}(T_{r,i})$ is equal to one of the following expressions in six different cases depending on the parameters (see Figure~\ref{fig:alg_cases} for illustration of the cases). 
	
	\[ 
	\begin{array}{ll}
		\displaystyle\min_{\substack{\ssc 0\le \omega_1 \le \pi(V(T_{v_1})) \\ \ssc \mathsf{b}_1\in\{\textsf{true},\textsf{false}\}}}
		\Bigl\{\bar{\psi}_{k,\omega_1,\omega_1,\rho-1,\mathsf{b}_1}(T_{v_1})\Bigr\} %
		& \text{ if } i=1 \text{ and } \omega=0 \\
		\displaystyle\min_{\substack{\ssc 0\le \omega_1 \le \pi(V(T_{v_1})) \\ \ssc \mathsf{b}_1\in\{\textsf{true},\textsf{false}\}}}
		\Bigl\{\bar{\psi}_{k-1,\omega_1,\omega_1,\rho,\mathsf{b}_1}(T_{v_1})\Bigr\} \textstyle + \frac{c(r\hat{r}) + c(rv_1)}{\mathsf{d}}    %
		& \text{ if } i=1 \text{ and } \omega>0 \text{ and } \mathsf{b}=\textsf{false} \\
		\displaystyle\min_{\substack{\ssc \mathsf{b}_1\in\{\textsf{true},\textsf{false}\}}}
		\Bigl\{\bar{\psi}_{k,\omega-\pi(v),\mathsf{d},\rho,\mathsf{b}_1}(T_{v_1})\Bigr\} \textstyle + \frac{c(r\hat{r}) - c(rv_1)}{\mathsf{d}} %
		& \text{ if } i=1 \text{ and } \omega>0 \text{ and } \mathsf{b}=\mathsf{true} \\
		\!\! \displaystyle\min_{\substack{\ssc 0\le k_1\le k \\ \ssc 0\le \rho_1 \le \rho \\ \ssc 0\le \omega_2 \le \pi(V(T_{v_i})) \\ \ssc \mathsf{b}_2\in\{\textsf{true},\textsf{false}\}}}%
		\!\! \Bigl\{\bar{\psi}_{k_1,0,0,\rho_1,\textsf{false}}(T_{r,i-1}) + \bar{\psi}_{k-k_1,\omega_2,\omega_2,\rho-\rho_1,\mathsf{b}_2}(T_{v_i}) \Bigr\} %
		& \text{ if } i>1 \text{ and } \omega=0 \\
		\!\! \displaystyle\min_{\substack{\ssc 0\le k_1\le k \\ \ssc 0\le \rho_1 \le \rho \\ \ssc 0\le \omega_2 \le \pi(V(T_{v_i})) \\ \ssc \mathsf{b}_1,\mathsf{b}_2\in\{\textsf{true},\textsf{false}\}}}%
		\!\! \Bigl\{\bar{\psi}_{k_1,\omega,\mathsf{d},\rho_1,\mathsf{b}_1}(T_{r,i-1}) + %
		\bar{\psi}_{k-k_1,\omega_2,\omega_2,\rho-\rho_1,\mathsf{b}_2}(T_{v_i}) \Bigr\} \textstyle + \frac{c(rv_i)}{\mathsf{d}}  %
		& \text{ if } i>1 \text{ and } \omega>0 \text{ and } \mathsf{b}=\textsf{false} \\
		\!\! \displaystyle\min_{\substack{\ssc 0\le k_1\le k \\ \ssc 0\le \rho_1 \le \rho \\ \ssc 0\le \omega_1 \le \omega \\ \ssc \mathsf{b}_1,\mathsf{b}_2\in\{\textsf{true},\textsf{false}\}}}%
		\!\! \Bigl\{\bar{\psi}_{k_1,\omega_1,\mathsf{d},\rho_1,\mathsf{b}_1}(T_{r,i-1}) + %
		\bar{\psi}_{k-k_1+1,\omega-\omega_1,\mathsf{d},\rho-\rho_1,\mathsf{b}_2}(T_{v_i})   \} \Bigr\} \textstyle - \frac{c(rv_i)}{\mathsf{d}}  %
		& \textstyle \text{ if } i>1 \text{ and } \omega>0 \text{ and } \mathsf{b}=\textsf{true} \\

	\end{array}
	\]
\end{theorem}

\begin{figure}[t]
	\label{fig:alg_cases}
	\begin{tikzpicture}[y=-1cm]
	\filldraw[black] (7.54889,2.39111) circle (0.05111cm);
	\definecolor{penColor}{gray}{0.8}
	\filldraw[penColor] (9.23333,2.39111) circle (0.05111cm);
	\draw (7.54889,2.85778) ellipse (0.63111cm and 0.71333cm);
	\filldraw[black] (7.96889,1.55111) circle (0.05111cm);
	\filldraw[penColor] (10.79111,6.68667) circle (0.05111cm);
	\filldraw[black] (9.23333,6.68667) circle (0.05111cm);
	\filldraw[black] (8.09556,6.68667) circle (0.05111cm);
	\filldraw[black] (6.75111,6.68667) circle (0.05111cm);
	\draw (9.23333,7.18889) ellipse (0.56889cm and 0.71333cm);
	\filldraw[black] (13.69333,2.39111) circle (0.05111cm);
	\filldraw[penColor] (15.37778,2.39111) circle (0.05111cm);
	\draw (13.69333,2.85778) ellipse (0.63111cm and 0.71333cm);
	\filldraw[black] (14.11778,1.55111) circle (0.05111cm);
	\fill[draw=black] (13.23111,5.84222) circle (0.05111cm);
	\filldraw[penColor] (16.85111,6.68667) circle (0.05111cm);
	\filldraw[black] (15.29556,6.68667) circle (0.05111cm);
	\filldraw[black] (14.15556,6.68667) circle (0.05111cm);
	\filldraw[black] (12.81111,6.68667) circle (0.05111cm);
	\draw (15.29556,7.18889) ellipse (0.56889cm and 0.71333cm);
	\fill[draw=black] (7.16889,5.84222) circle (0.05111cm);
	\filldraw[black] (1.65778,2.39111) circle (0.05111cm);
	\filldraw[penColor] (3.34222,2.39111) circle (0.05111cm);
	\draw (1.65778,2.85778) ellipse (0.63111cm and 0.71333cm);
	\draw[black] (2.07778,1.55111) circle (0.05111cm);
	\draw[black] (1.31778,5.84222) circle (0.05111cm);
	\filldraw[penColor] (4.93778,6.68667) circle (0.05111cm);
	\filldraw[black] (3.38222,6.68667) circle (0.05111cm);
	\filldraw[black] (2.24444,6.68667) circle (0.05111cm);
	\filldraw[black] (0.9,6.68667) circle (0.05111cm);
	\draw (3.38222,7.18889) ellipse (0.56889cm and 0.71333cm);
	\draw (7.96444,1.55111) circle (0.27333cm);
	\draw (14.108667,1.54) circle (0.27333cm);
	\draw[black] (7.54889,2.26667) -- (7.13111,3.23778) -- (7.96889,3.23778) -- (7.54889,2.26667) -- cycle;
	\draw[penColor] (9.23333,2.26667) -- (8.81333,3.23778) -- (9.65333,3.23778) -- (9.23333,2.26667) -- cycle;
	\draw[penColor] (7.96889,1.55111) -- (9.23333,2.39111);
	\draw[dashed] (7.96889,1.55111) -- (7.54889,2.39111);
	\draw[penColor] (10.79111,6.56) -- (10.37111,7.52667) -- (11.21333,7.52667) -- (10.79111,6.56) -- cycle;
	\draw[black] (9.23333,6.56) -- (8.81333,7.52667) -- (9.65333,7.52667) -- (9.23333,6.56) -- cycle;
	\draw[black] (8.09556,6.56) -- (7.67556,7.52667) -- (8.51778,7.52667) -- (8.09556,6.56) -- cycle;
	\draw[black] (6.75111,6.56) -- (6.32889,7.52667) -- (7.16889,7.52667) -- (6.75111,6.56) -- cycle;
	\draw[dashed] (7.16889,5.84222) -- (9.23333,6.68667);
	\draw (7.16889,5.84222) -- (8.09556,6.68667);
	\draw[penColor] (7.16889,5.84222) -- (10.79111,6.68667);
	\draw[black] (13.69333,2.26667) -- (13.27333,3.23778) -- (14.11778,3.23778) -- (13.69333,2.26667) -- cycle;
	\draw[penColor] (15.37778,2.26667) -- (14.95556,3.23778) -- (15.8,3.23778) -- (15.37778,2.26667) -- cycle;
	\draw[penColor] (14.11778,1.55111) -- (15.37778,2.39111);
	\draw (14.11778,1.55111) -- (13.69333,2.39111);
	\draw (13.23111,5.84222) -- (12.81111,6.68667);
	\draw[penColor] (16.85111,6.56) -- (16.43111,7.52667) -- (17.27333,7.52667) -- (16.85111,6.56) -- cycle;
	\draw[black] (15.29556,6.56) -- (14.87111,7.52667) -- (15.71556,7.52667) -- (15.29556,6.56) -- cycle;
	\draw[black] (14.15556,6.56) -- (13.73778,7.52667) -- (14.57778,7.52667) -- (14.15556,6.56) -- cycle;
	\draw[black] (12.81111,6.56) -- (12.38667,7.52667) -- (13.23111,7.52667) -- (12.81111,6.56) -- cycle;
	\draw (13.23111,5.84222) -- (15.29556,6.68667);
	\draw (13.23111,5.84222) -- (14.15556,6.68667);
	\draw[penColor] (13.23111,5.84222) -- (16.85111,6.68667);
	\draw[black] (1.65778,2.26667) -- (1.23556,3.23778) -- (2.07778,3.23778) -- (1.65778,2.26667) -- cycle;
	\draw[penColor] (3.34222,2.26667) -- (2.92222,3.23778) -- (3.76,3.23778) -- (3.34222,2.26667) -- cycle;
	\draw[penColor] (2.07778,1.55111) -- (3.34222,2.39111);
	\draw[dashed] (2.07778,1.55111) -- (1.65778,2.39111);
	\draw[penColor] (4.93778,6.56) -- (4.52,7.52667) -- (5.36222,7.52667) -- (4.93778,6.56) -- cycle;
	\draw[black] (3.38222,6.56) -- (2.96222,7.52667) -- (3.80444,7.52667) -- (3.38222,6.56) -- cycle;
	\draw[black] (2.24444,6.56) -- (1.82444,7.52667) -- (2.66667,7.52667) -- (2.24444,6.56) -- cycle;
	\draw[black] (0.9,6.56) -- (0.47778,7.52667) -- (1.31778,7.52667) -- (0.9,6.56) -- cycle;
	\draw[dashed] (1.31778,5.84222) -- (3.38222,6.68667);
	\draw[penColor] (1.31778,5.84222) -- (4.93778,6.68667);
	\draw plot [smooth] coordinates { (1.3,5.7)  (2.4,6.65111)  (2.67111,7.6)  (0.43111,7.6)  (0.67111,6.7)  (1.3,5.7) };
	\draw plot [smooth] coordinates {(7.15,5.7)  (8.25,6.65111)  (8.52111,7.6)  (6.28111,7.6)  (6.52111,6.7)  (7.15,5.7)};
	\draw plot [smooth] coordinates {(13.2,5.7)  (14.3,6.65111)  (14.57111,7.6)  (12.33111,7.6)  (12.57111,6.7)  (13.2,5.7)};
	\draw (7.16889,5.84222) -- (6.75111,6.68667);
	\draw (1.31778,5.84222) -- (2.24444,6.68667);
	\draw (1.31778,5.84222) -- (0.9,6.68667);
	\path (8.14,2.39111) node[text=penColor,anchor=base west] {\tiny{}$\cdots$};
	\path (7.78667,1.76222) node[text=black,anchor=base west] {\tiny{}$r$};
	\path (7.00222,6.68667) node[text=black,anchor=base west] {\tiny{}$\cdots$};
	\path (9.73556,6.68667) node[text=penColor,anchor=base west] {\tiny{}$\cdots$};
	\path (14.28222,2.39111) node[text=penColor,anchor=base west] {\tiny{}$\cdots$};
	\path (13.93111,1.76222) node[text=black,anchor=base west] {\tiny{}$r$};
	\path (13.04667,6.05333) node[text=black,anchor=base west] {\tiny{}$r$};
	\path (13.06222,6.68667) node[text=black,anchor=base west] {\tiny{}$\cdots$};
	\path (15.8,6.68667) node[text=penColor,anchor=base west] {\tiny{}$\cdots$};
	\path (2.24444,2.39111) node[text=penColor,anchor=base west] {\tiny{}$\cdots$};
	\path (13.2,8.1) node[anchor=base west] {\fontsize{6.0}{7.2}\selectfont{}Case 6: $i>1, \Omega>0, t=\text{true}$};
	\path (7,8.1) node[anchor=base west] {\fontsize{6.0}{7.2}\selectfont{}Case 5: $i>1, \Omega>0, t=\text{false}$};
	\path (1.15333,6.68667) node[text=black,anchor=base west] {\tiny{}$\cdots$};
	\path (3.88667,6.68667) node[text=penColor,anchor=base west] {\tiny{}$\cdots$};
	\path (1.2,8.1) node[anchor=base west] {\fontsize{6.0}{7.2}\selectfont{}Case 4: $i>1, \Omega=0$};
	\path (7,3.9) node[anchor=base west] {\fontsize{6.0}{7.2}\selectfont{}Case 2: $i=1, \Omega>0, t=\text{false}$};
	\path (1.3,3.9) node[anchor=base west] {\fontsize{6.0}{7.2}\selectfont{}Case 1: $i=1, \Omega=0$};
	\path (13.1,3.9) node[anchor=base west] {\fontsize{6.0}{7.2}\selectfont{}Case 3: $i=1, \Omega>0, t=\text{true}$};
	\path (9,6.95) node[text=black,anchor=base west] {\tiny{}$v_i$};
	\path (7.8,6.95) node[text=black,anchor=base west] {\tiny{}$v_{i\!-\!1}$};
	\path (6.5,6.95) node[text=black,anchor=base west] {\tiny{}$v_1$};
	\path (12.55,6.95) node[text=black,anchor=base west] {\tiny{}$v_1$};
	\path (13.9,6.95) node[text=black,anchor=base west] {\tiny{}$v_{i\!-\!1}$};
	\path (15.05,6.95) node[text=black,anchor=base west] {\tiny{}$v_i$};
	\path (13.45,2.65) node[text=black,anchor=base west] {\tiny{}$v_1$};
	\path (7.3,2.65) node[text=black,anchor=base west] {\tiny{}$v_1$};
	\path (1.4,2.65) node[text=black,anchor=base west] {\tiny{}$v_1$};
	\path (3.12,7) node[text=black,anchor=base west] {\tiny{}$v_i$};
	\path (1.95,7) node[text=black,anchor=base west] {\tiny{}$v_{i\!-\!1}$};
	\path (0.63,7) node[text=black,anchor=base west] {\tiny{}$v_1$};
	\path (1.9,1.8) node[text=black,anchor=base west] {\tiny{}$r$};
	\path (7,6.1) node[text=black,anchor=base west] {\tiny{}$r$};
	\path (1.14,6.1) node[text=black,anchor=base west] {\tiny{}$r$};

	\end{tikzpicture}%
	\caption{The six cases of Theorem~\ref{thm:tree}.
		Parts of the tree that are not considered in the current subproblem are shown in gray. The vertex $r$ is hollow when it is an outlier, i.e.\ $\omega=0$. 
		The edge $rv_i$ is dashed when it should not be inside a part, i.e.\ $\mathsf{b} = \textsf{false}$.
	}
\end{figure}

\begin{proof}
	
	We discuss each case in turn. 
	\begin{description}
		\item[$i=1$]
		In this case the tree under consideration contains only the first child of $r$ (and its associated subtree). 
		We have two cases based on the value of $\omega$. 
		\begin{description}
			\item[$\omega=0$] [Case 1]
			This means that only subpartitions in which $r$ falls into the outlier set  are to be considered. 
			Hence, if $\mathcal{A}\in \mathcal{CD}_{k,0,\rho,\textsf{false}}(T_{r,1})$, then $\mathcal{A}\in \mathcal{CD}_{k,\omega_1,\rho-1,\mathsf{b}_1}(T_{v_1})$
			for some parameters $\omega_1$ and $\mathsf{b}_1$. Moreover, 
			
			\begin{equation}
				\label{eqn:AA'1}	
				\Psi^{\mathsf{d}}_{T_{r,1}}(\mathcal{A}) = \Psi_{T_{r,1}}(\mathcal{A}) = \Psi^{\omega_1}_{T_{v_1}}(\mathcal{A}).
			\end{equation} 
			
			The converse also holds; i.e.\ each 
			subpartition $\mathcal{A}'\in \mathcal{CD}_{k,\omega_1,\rho-1,\mathsf{b}_1}(T_{v_1})$ for some parameters $\omega_1, \mathsf{b}_1$ is automatically
			a member of $\mathcal{CD}_{k,0,\rho,\mathsf{false}}(T_{r,1})$ with the same $\Psi$ value.  
			This means that an optimal subpartition for $T_{r,1}$ with parameters $k,0, \mathsf{d}, \rho, \textsf{false}$ corresponds to an optimal 
			subpartition for $T_{v_1}$ with parameters $k, \omega_1, \mathsf{d}'=\omega_1, \rho-1$ and $\mathsf{b}_1$ for some $\omega_1$ and $\mathsf{b}_1$. 
			Hence, the conclusion in this case is settled. 
			
			\item[$\omega\ge 1$] 
			This means that $r$ is not an outlier. We distinguish two cases based on whether $r$ is in the same part as $v_i$
			or not. 
			\begin{description}
				\item[$\mathsf{b}=\mathsf{false}$] [Case 2]
					As $i=1$, only subpartitions for which $r$ is the only vertex in its part are to be considered. Hence, we may assume $\omega = \pi(r)$
				(otherwise $\mathcal{CD}_{k,\omega,\rho,\mathsf{b}}(T_{r,1})=\emptyset$). 
				Now, if we take any $\mathcal{A}\in \mathcal{CD}_{k,\omega,\rho,\mathsf{b}}(T_{r,1})$ and remove the part containing $r$ to get 
				$\mathcal{A}'$, then $\mathcal{A}'\in \mathcal{CD}_{k-1,\omega_1,\rho,\mathsf{b}_1}(T_{v_1})$ for some $\omega_1$ and $\mathsf{b}_1$. 
				Furthermore, 
				
				\begin{equation}
					\label{eqn:AA'2}	
					\Psi^{\mathsf{d}}_{T_{r,1}}(\mathcal{A}) = \frac{c(r\hat{r}) + c(rv_1)}{\mathsf{d}} + \Psi^{\mathsf{d}}_{T_{v_1}}(\mathcal{A}').
				\end{equation} 
				
				In a similar manner to the previous case, the converse also holds, i.e.\ by adding a part consisting of only $r$ to any such $\mathcal{A}'$,
				one can get the corresponding subpartition $\mathcal{A}$. 
				\item[$\mathsf{b}=\mathsf{true}$] [Case 3]
					In this case the vertices $r$ and $v_1$ are in the same part (and hence we assume $\omega \ge \pi(r)+\pi(v_1)$). 
				Take any $\mathcal{A}\in \mathcal{CD}_{k,\omega,\rho,\mathsf{b}}(T_{r,1})$, and suppose $r\in A_l \in \mathcal{A}$. 
					Let $\mathcal{A}'$ be the same as $\mathcal{A}$ except that we replace $A_l$ by $A_l\setminus \{r\}$. 
					Then $\mathcal{A}'\in \mathcal{CD}_{k,\omega-\pi(r),\rho,\mathsf{b}_1}(T_{v_1})$ for some $\mathsf{b}_1$. 
				Furthermore, 
				\begin{equation}
					\label{eqn:AA'3}	
					\Psi^{\mathsf{d}}_{T_{r,1}}(\mathcal{A}) = \frac{c(r\hat{r})}{\mathsf{d}} + \Psi^{\mathsf{d}}_{T_{v_1}}(\mathcal{A}') - \frac{c(rv_1)}{\mathsf{d}}.
				\end{equation} 
					Again, the converse also holds; that is, for any $\mathcal{A}'\in \mathcal{CD}_{k,\omega-\pi(r),\rho,\mathsf{b}_1}(T_{v_1})$, we can add $r$ to 
				the part containing $v_1$ to get a subpartition $\mathcal{A}\in \mathcal{CD}_{k,\omega,\rho,\mathsf{b}}(T_{r,1})$, where \ref{eqn:AA'3} holds.
			\end{description}
		\end{description}
		\item[$i\ge 2$]
		There are two cases based on the value of $\omega$. 
		\begin{description}
			\item[$\omega=0$] [Case 4]
			If $\mathcal{A}\in \mathcal{CD}_{k,0,\rho,\mathsf{b}}(T_{r,i})$, then $\mathcal{A}$ is a union of two subpartitions for $T_{r,i-1}$ and $T_{v_i}$; i.e.\ 
			$\mathcal{A} = \mathcal{A}' \cup \mathcal{A}''$ where 
				$\mathcal{A}'\in \mathcal{CD}_{k_1,0,\rho_1,\mathsf{false}}(T_{r,i-1})$ and $\mathcal{A}''\in \mathcal{CD}_{k-k_1,\omega_2,\rho-\rho_1,\mathsf{b}_2}(T_{v_i})$ 
			for some parameters $k_1, \rho_1, \omega_2$ and $\mathsf{b}_2$. Furthermore, we have
			\begin{equation}
				\label{eqn:AA'4}	
				\Psi^{\mathsf{d}}_{T_{r,i}}(\mathcal{A}) = \Psi_{T_{r,i}}(\mathcal{A}) = \Psi_{T_{r,i-1}}(\mathcal{A}') + \Psi^{{\omega_2}}_{T_{v_i}}(\mathcal{A}'').
			\end{equation} 
			Again, such subpartitions $\mathcal{A}'$ and $\mathcal{A}''$ may be combined to give a subpartition $\mathcal{A}\in \mathcal{CD}_{k,0,\rho,\mathsf{b}}(T_{r,i})$
			for which the same relation \ref{eqn:AA'2} still holds. 
			
			\item[$\omega \ge 1$] 
			Here again we consider the two possible values for $\mathsf{b}$. 
			\begin{description}
				\item[$\mathsf{b}=\mathsf{false}$] [Case 5]
				If $\mathcal{A}\in \mathcal{CD}_{k,\omega,\rho,\textsf{false}}(T_{r,i})$, then as $\mathsf{b}=\textsf{false}$, 
				the subpartition $\mathcal{A}$ is a union of two subpartitions for $T_{r,i-1}$ and $T_{v_i}$; i.e.\ 
				$\mathcal{A} = \mathcal{A}' \cup \mathcal{A}''$ where 
				$\mathcal{A}'\in \mathcal{CD}_{k_1,\omega,\rho_1,\mathsf{b}_1}(T_{r,i-1})$ and $\mathcal{A}''\in \mathcal{CD}_{k-k_1,\omega_2,\rho-\rho_1,\mathsf{b}_2}(T_{v_i})$ 
				for some parameters $k_1, \rho_1, \omega_2, \mathsf{b}_1$ and $\mathsf{b}_2$. Furthermore, we have
				\begin{equation}
					\label{eqn:AA'5}	
					\Psi^{\mathsf{d}}_{T_{r,i}}(\mathcal{A}) = \Psi^{\mathsf{d}}_{T_{r,i-1}}(\mathcal{A}') + \frac{c(rv_i)}{\mathsf{d}} + \Psi^{{\omega_2}}_{T_{v_i}}(\mathcal{A}'') 
				\end{equation} 
				Here, also such subpartitions $\mathcal{A}'$ and $\mathcal{A}''$ may be combined to give $\mathcal{A}\in \mathcal{CD}_{k,0,\rho,\mathsf{b}}(T_{r,i})$
				with the same relation \ref{eqn:AA'3} holding. 
				
				\item[$\mathsf{b}=\mathsf{true}$] [Case 6]
					In this case vertices $r$ and $v_i$ need to be in the same part, and hence, we may assume $\omega \ge \pi(r)+\pi(v_i)$. 
				Take any subpartition $\mathcal{A}\in \mathcal{CD}_{k,\omega,\rho,\mathsf{b}}(T_{r,i})$, and suppose $r\in A_l \in \mathcal{A}$. 
				Then $A_l$ may be partitioned into $A_l'\subseteq T_{r,i-1}$ and $A_l''\subseteq T_{v_i}$. 
				
				Let 
				$$ \mathcal{A}' \isdef \{S\in \mathcal{A} \mid S\subseteq T_{r,i-1} \} \cup \{ A_l' \} $$
				and 
				$$ \mathcal{A}'' \isdef  \{S\in \mathcal{A} \mid S\subseteq T_{v_i} \} \cup \{ A_l'' \}. $$
				
				Then, we have $$\mathcal{A}'\in \mathcal{CD}_{k_1,\omega_1,\rho_1,\mathsf{b}_1}(T_{r,i-1})$$ for some $k_1, \omega_1, \rho_1, \mathsf{b}_1$, and 
				$$\mathcal{A}''\in \mathcal{CD}_{k-k_1+1,\omega-\omega_1,\rho-\rho_1,\mathsf{b}_2}(T_{v_i})$$ for some $\mathsf{b}_2$, while we have 
				\begin{equation}
            	\label{eqn:AA'6}	
             	\Psi^{\mathsf{d}}_{T_{r,i}}(\mathcal{A}) = \Psi^{\mathsf{d}}_{T_{r,i-1}}(\mathcal{A}') + \Psi^{\mathsf{d}}_{T_{v_i}}(\mathcal{A}'') - \frac{c(rv_i)}{\mathsf{d}}
               \end{equation} 
				
				The converse also holds; that is, by combining any such subpartitions $\mathcal{A}'$ and $\mathcal{A}''$ one may get a subpartition  
				$\mathcal{A}\in \mathcal{CD}_{k,\omega,\rho,\mathsf{b}}(T_{r,i})$ for which the same equation holds.
				
			\end{description}
			
		\end{description}
		
	\end{description}
	
\end{proof}

As remarked above, Theorem~\ref{thm:tree} can readily be turned into a dynamic programming algorithm that solves $\mathsf{CRISO}^{1}(T,k,\rho)$ (see Algorithm~\ref{alg:iso} for the pseudocode).
This leads to our first main result. 

\begin{theorem}
	\label{thm:iso}
	For any given weighted tree $T$ the problems $\mathsf{CRISO}^{1}(T,k,\rho)$, $\mathsf{CISO}^{1}(T,k)$, and $\mathsf{ISO}^{1}(T,k)$
      are solvable in time $O(k^2 \rho^2\ \pi(V(T)^3)$.  In particular, all problems are polynomial-time solvable when $T$ has polynomially bounded vertex-weights.

\end{theorem}
\begin{proof} 
	First, let us present a running time analysis of the algorithm which is based on Theorem~\ref{thm:tree}.
	Note that the possible range of parameters $k,\omega,\mathsf{d},\rho, \mathsf{b}$, yields a total of 
	$O(k \rho \pi(V(T_{r_0}))^2)$ subproblems for a given tree $T_{r_0}$. Also, by considering the six cases of Theorem~\ref{thm:tree}, one may verify  
	that each $\bar{\psi}_{k,\omega,\mathsf{d},\rho,\mathsf{b}}(T_{v,i})$ 
	can be written as the minimum of $O(k \rho \pi(V(T_{r_0})))$ expressions, each of which involving at most two smaller subproblems. 
	Hence, the running time of Algorithm~\ref{alg:iso} is $O(k^2 \rho^2 \pi(V(T_{r_0}))^3)$. 
	For polynomially bounded $\pi$, this is polynomial.

	By Proposition~\ref{prop:psilongpsi}, in order to solve $\mathsf{CRISO}^{1}(T,k,\rho)$, it suffices to compute $\bar{\psi}_{k,\omega,\omega,\rho,\mathsf{b}}(T_{r_0})$ 
	for all possible values of $\omega$, and $\mathsf{b}$ and take the best solution. Theorem~\ref{thm:tree} guarantees that this can be done in time $O(k^2 \rho^2\ \pi(V(T)^3)$.  
	This also shows the solvability of $\mathsf{CISO}^{1}(T,k)$ in the same running time, since we have $\bar{\psi}_{k}^1(G) = \bar{\psi}_{k,n}^1(G)$ for any weighted graph $G$.
	The solvability of $\mathsf{ISO}^{1}(T,k)$ with the same running time follows from the solvability of $\mathsf{CISO}^{1}(T,k)$ and Theorem~\ref{thm:daneshgarjavadi-connected}. 
\end{proof}



\SetAlFnt{\scriptsize}	
\begin{algorithm}[h] 
	\SetKwData{False}{false}\SetKwData{True}{true}
	\SetKwData{ia}{$k$}\SetKwData{ib}{numOutliers}\SetKwData{ic}{isOut}
	\SetKwData{id}{$\mathsf{d}$}\SetKwData{ie}{$\omega$}\SetKwData{ig}{$\mathsf{b}$}
	\SetKwData{iaa}{$k_1^\ast$}\SetKwData{ibb}{numOutliers1}\SetKwData{icc}{isOut'}
	\SetKwData{idd}{$C_1$}\SetKwData{iee}{$\omega_1^\ast$}\SetKwData{igg}{$\mathsf{b}_1^\ast$}
	\SetKwData{iaaa}{$k_2^\ast$}\SetKwData{ibbb}{numOutliers2}\SetKwData{iccc}{isOut'{'}}
	\SetKwData{iddd}{$C_2$}\SetKwData{ieee}{$\omega_2^\ast$}\SetKwData{iggg}{$\mathsf{b}_2^\ast$}
	\SetKwData{io}{$\rho$}\SetKwData{ioo}{$\rho_1^\ast$}
	\SetKwData{END}{lastChild}
	\SetKw{FOR}{For}
	\SetKw{ASSERT}{assert}
	\SetKw{Fn}{function}
	\SetKwFunction{iso}{iso} \SetKwFunction{isoo}{mainIso}
	\SetKwInOut{Input}{input}\SetKwInOut{Output}{output}
	
	\Fn{\iso{$r$, $i$, \ia, \ie, \id, \io, \ig }}{ \\
		\tcp{returns $\bar{\psi}_{k,\omega,\mathsf{d},\rho,\mathsf{b}}(T_{r,i})$}
		\tcp{for brevity, the cases where a value of $\infty$ has to be returned have been left out of code and have been mentioned in the comments}
		\tcp{note that an asterisk superscript indicates that the minimization is over all possible values for the corresponding parameter; %
			see Theorem~\ref{thm:tree} for the exact range of parameters} 
		\tcp{each of the 6 cases other than the base case have been numbered with the corresponding case number in Theorem \ref{thm:tree}}
		\Begin{
			\uIf(\tcp*[h]{base case. we must have $k,\ie \le 1$ and $\io = 1-\ie$.}){$r$ is a leaf}{ 
				\eIf{$\ie=0$}{ %
					\Return 0
				}
				{
					\Return $\frac{c(r\hat{r})}{\mathsf{d}}$
				}
				
			}
			\uElseIf (\tcp*[h]{$T'$ contains only the first child of $r$}) {$i=1$ }{ 
				\eIf(\tcp*[h]{$r$ is an outlier \color{red}{[Case 1]}}) {$\ie = 0$}{ 
					\Return $\min
					\iso(v_1,\END(v_1),\ia,\iee,\iee,\io-1,\igg)$ \nllabel{min_1} 
				}
				(// $r$ is not an outlier){ 
					\eIf(\tcp*[h]{$r$ is in a part by itself, hence we must have $\ie=\pi(r)$ \color{red}{[Case 2]}}){\ig = \False}{
						\Return $\frac{c(r \hat{r})+c(rv_1)}{\mathsf{d}}$+ $\min
						\iso(v_1,\END(v_1),\ia-1,\iee,\iee,\io,\igg)$
					}
					(\tcp*[h]{$r$ is in the same part as $v_1$, hence we must have $\ie \ge \pi(r)+\pi(v_1)$ \color{red}{[Case 3]}}){ 
						\Return $\frac{c(r \hat{r})-c(rv_1)}{\mathsf{d}}$+ $\min
						\iso(v_1,\END(v_1),\ia,\ie-\pi(r),\id,\io,\igg)$ 
					}
					
				}
				
			}
			\uElse(\tcp*[h]{$i\ge 2$}) {  
				\eIf(\tcp*[h]{$r$ is an outlier \color{red}{[Case 4]}}){$\ie = 0$}{
					\Return $\min
					(\iso(r,i-1, \iaa,0,0,\ioo,\False) + 
					\iso(v_i,\END(v_i),\ia-\iaa,\ieee,\ieee,\io-\ioo,\iggg))$
				}
				(\tcp*[h]{$r$ is not an outlier}){ 
					\eIf(\tcp*[h]{$r$ is not in the same part as $v_i$ \color{red}{[Case 5]}}){\ig = \False}{
						\Return $\frac{c(rv_i)}{\mathsf{d}}$+$\min%
						(\iso(r,i-1,\iaa,\ie,\id,\ioo,\igg) + 
						\iso(v_i,\END(v_i),\ia-\iaa,\ieee,\ieee,\io-\ioo,\iggg))$ %
					}
					(\tcp*[h]{$r$ is in the same part as $v_i$, hence we must have $\ie \ge \pi(r)+\pi(v_i)$ \color{red}{[Case 6]}}){ 
						\Return $\frac{-c(rv_i)}{\mathsf{d}}$+$\min
						(\iso(r,i-1,\iaa,\iee,\id,\ioo,\igg)+
						\iso(v_i,\END(v_i),\ia-\iaa+1,\ie-\iee,\id,\io-\ioo,\iggg))$, \qquad \qquad
						where  $0 < \iee < \ie $. \nllabel{cons2}
					}
				}
			}
		}
	}
	
	\caption{Mean isoperimetry algorithm on a weighted tree.}
	\label{alg:iso}
\end{algorithm}

We shall see in Section~\ref{sec:general} that $\mathsf{RISO}^1$ can be 2-approximated on weighted trees; see Corollary~\ref{cor:2alpha}. 
\begin{remark}
	\label{rem:isooptimizer}
	In a similar way to most dynamic programming algorithms,
	by keeping more information in each cell of the global table, Algorithm~\ref{alg:iso} may be modified to 
	compute the optimal subpartition itself. 
	For example, suppose
	\[ 
	\bar{\psi}_{k,\omega,\mathsf{d},\rho,\mathsf{b}}(T_{r,i}) = 
	C + \psi_{k_1,\omega_1,\mathsf{d}_1,\rho_1,\mathsf{b}_1}(T_{r_1,i_1}) +  \psi_{k_2,\omega_2,\mathsf{d}_2,\rho_2,\mathsf{b}_2}(T_{r_2,i_2}) 
	\]
	for some $C$. 
	Then in the cell corresponding to $(r,i,k,\mathsf{d},\omega,\mathsf{b})$ we keep 
	$$(r_1,i_1,k_1,\mathsf{d}_1,\omega_1,\mathsf{b}_1,r_2,i_2,k_2,\mathsf{d}_2,\omega_2,\mathsf{b}_2).$$
	The optimal subpartition for the subproblem with parameters $(r,i,k,\mathsf{d},\omega,\mathsf{b})$ can then be constructed by 
	recursively building optimal solutions for subproblems 
	$$(r_1,i_1,k_1,\mathsf{d}_1,\omega_1,\mathsf{b}_1) \ \  {\rm and} \ \ (r_2,i_2,k_2,\mathsf{d}_2,\omega_2,\mathsf{b}_2)$$
	and combining them. 
\end{remark}

\begin{remark}
	\label{rem:crisomax}
	The same general idea used in Theorem~\ref{thm:tree} can be used to develop polynomial-time algorithms for other versions of the isoperimetry problem. 
	Here we mention a few such cases. 
	\begin{itemize}

		\item \textbf{Solving $\mathsf{CRISO}^\infty$ on weighted trees} \\
			For this one may define the parameter $\bar{\psi}^\infty_{k,\omega,\mathsf{d},\rho,\mathsf{b}}(T_{r}, \eta)$
			to indicate the minimum of $\frac{c(\delta(A_k))}{d}$ where the minimization is taken over all subpartitions $\mathcal{A} = (A_1, A_2, \ldots, A_k)$ of $V(T_r)$ 
			with $r\in A_k$ and $\frac{c(\delta(A_i))}{\pi(A_i)} \le \eta$, for all $1\le i \le k-1$. 
			Then, in the same vein as in Theorem~\ref{thm:tree},  one may prove a similar result characterizing 
			$\bar{\psi}^\infty_{k,\omega,\mathsf{d},\rho,\mathsf{b}}(T_{r}, \eta)$ as a sum of similar parameters on subtrees of $T_r$.
			This gives an algorithm for computing the \textit{decision version},  
			$\mathsf{DCRISO}^\infty$, on weighted trees. 
			One may then use binary search to find the actual optimal value. 

		\item \textbf{Restricting outliers to a fixed set} \\ 
			It might be desirable in certain applications to have the constraint that all of the, at most $\rho$, outliers must 
			reside in a (possibly proper) subset $U$ of the vertex set of the graph. 
			This can be handled by generalizing $\bar{\psi}_{k,\omega,\mathsf{d},\rho,\mathsf{b}}(T_{r,i})$ to 
			$\bar{\psi}_{k,\omega,\mathsf{d},\rho,\mathsf{b}, U}(T_{r,i})$ which explicitly stipulates a set $U\subseteq V(T_{r,i})$ 
			in which the outliers are permitted to lie. 

		\item \textbf{Potentials on vertices} \\
			All the versions mentioned above can be generalized to the setting where a \textit{potential} function
			$p:V \to \mathbb{Z}$ is given on the vertex set, in which case the normalized cut value for 
			a set $U\subseteq V$ is defined to be $\frac{c(\delta(U))+p(U)}{w(U)}$. The potential $p(v)$ can be thought of as the 
			weight of an edge between $v$ and a hypothetical \textit{ground} vertex $v_0$ which is necessarily an outlier.

	\end{itemize}

\end{remark}

\section{Approximation algorithms}
\label{sec:general}

In this section, we consider approximating all the previously discussed variants of the mean isoperimetry problem. 
It is quite interesting to note that although these problems behave differently when 
seeking exact algorithms, it will turn out that all variants of the mean isoperimetry problem are computationally equivalent up to 
constant factor approximation (i.e.\ an $\alpha$-approximation for one of them yields an $8\alpha$-approximation for any other version).


Our technique on the inequality side relies on some well-known inequalities and facts (see Proposition~\ref{pro:basicineq}) for the ordinary isoperimetry problem as well as a new result of ours 
showing the approximability of the connected versions (see Proposition~\ref{pro:connected}). On the other hand, and on the computational side, our main result relies on an approximation of 
$\mathsf{ISO}^{1}(G,k)$ for general weighted graphs. For this, we will apply our polynomial-time algorithm of the previous section to the R{\"a}cke-Shah tree cut sparsifier 
with suitable vertex weights, and we will show that this leads to an approximation for the original problem. 
We will also address the problem of finding the (approximately) optimal subpartitions (as opposed to approximating only the objective function), 
and will argue that this problem can also be solved using our approach. 


To begin, let us recall a couple of facts and put them together in the following proposition for future reference.
\begin{proposition}\label{pro:basicineq}
	The following inequalities hold for different variants of the isoperimetry problem on a connected weighted graph $G$. 
	\begin{itemize}	
		
		\item[{\rm a)}] $\rho \le \rho' \ \ \Rightarrow \ \ \psi^p_{k,\rho'}(G) \le \psi^p_{k,\rho}(G)$.
		\item[{\rm b)}] $\rho \le \rho' \ \ \Rightarrow \ \ \bar{\psi}^p_{k,\rho'}(G) \le \bar{\psi}^p_{k,\rho}(G)$.
		\item[{\rm c)}] $\psi_{2,0}^p(G)=\psi_{2,n}^p(G)=\psi_{2}^p(G)$.	
		\item[{\rm d)}] $\bar{\psi}_{k}^p(G)=\bar{\psi}_{k,n}^p(G)=\psi_{k,n}^p(G)=\psi_{k}^p(G)$.
		\item[{\rm e)}] $\psi_{k,0}^1(G) < 2 \psi_{k,n}^1(G) = 2 \psi_{k}^1(G).$
		\item[{\rm f)}] $\psi_{k,0}^\infty(G) < (k-1) \psi_{k,n}^\infty(G) = (k-1) \psi_{k}^\infty(G).$
	\end{itemize}	
	
\end{proposition}
\begin{proof}	Parts (a) and (b) are clear by definitions. For Part (c) see \cite{daneshgar-javadi-jctb} Proposition~1, and for Part (d) see \cite{daneshgar-javadi-trees} Lemma~2.
	Parts (e) and (f) are results of Theorem~1 of \cite{daneshgar-javadi-trees}. 
\end{proof}	

As a direct corollary of Proposition~\ref{pro:basicineq}, one may conclude that the robust version of the mean isoperimetry problem 
is essentially equivalent to $\mathsf{ISO}^{1}$ up to constant factors. More precisely, 

\begin{corollary}
	\label{cor:2alpha}
	A factor $\alpha$ polynomial approximation algorithm for $\mathsf{ISO}^1$ on a class of graphs yields a factor $2\alpha$ polynomial approximation algorithm for 
	$\mathsf{RISO}^1$ on the same class of graphs.
\end{corollary}
\begin{proof}
	By parts (a) and (e) of Proposition~\ref{pro:basicineq}, 
	\[
	\psi_{k}^1(G)=\psi_{k,n}^1(G) \le \psi_{k,\rho}^1(G) \le \psi_{k,0}^1(G) < 2 \psi_{k,n}^1(G)=2 \psi_{k}^1(G).
	\]
	Hence, in order to approximate $\psi_{k,\rho}^1(G)$, 
	it suffices to compute $\psi_{k,n}^1$ (exactly or approximately), and output twice its value. 
\end{proof}

The case of the connected variants of the isoperimetry is more subtle. 
In the following proposition, we show that approximating the connected case is also equivalent to approximating $\mathsf{ISO}^{1}(G,k)$ up to a constant factor equal to $8$.

\begin{proposition}
	\label{pro:connected}
	For a connected weighted graph $G$ the following inequalities hold,	
	\begin{enumerate}
		\item[{\rm a)}] $\bar{\psi}_{k,0}^1(G) \le 4\psi_{k,0}^1(G)$. 
		\item[{\rm b)}]	$\bar{\psi}_{k,\rho}^1(G) \le 8\psi_{k,n}^1(G) \le 8\psi_{k,\rho}^1(G). $
	\end{enumerate}
\end{proposition}

\begin{proof}
	First let us define $\mathcal{D}'_{k,0}(V(G))$ as the set of all $k$-partitions  $\mathcal{A} = (A_1, \ldots, A_k)$ of $V(G)$ where for each $2\le i\le k$, the graph $G[A_i]$ is connected (i.e.\  only one part $A_1$ may induce a disconnected subgraph). Also, let 
	$$\mathcal{B} \isdef \mathrm{arg}\min \{ \Psi_G(\mathcal{A}) \mid \mathcal{A}\in \mathcal{D}'_{k,0}(V(G)) \}.$$
	In what follows we show that 
	$$  \frac{1}{2} \bar{\psi}_{k,0}^1(G) \le  \Psi_G(\mathcal{B}) \le 2\psi_{k,0}^1(G). $$
	\begin{itemize}
		\item First, let us prove $\Psi_G(\mathcal{B}) \le 2\psi_{k,0}^1(G)$. For this, 
		let $\mathcal{A} = (A_1, \ldots, A_k)$ be a $k$-partition of $V(G)$ minimizing the mean isoperimetry objective function, that is
		$$ \psi_{k,0}^1(G) = \Psi_G(\mathcal{A}).$$
		Furthermore, assume $\pi(A_1) \ge \pi(A_i)$ for all $2\le i\le k$. 
		The main idea is, for each $2\le i\le k$, to keep the best component of $A_i$, and add the other components to $A_1$. 
		
		Let $A^1_i, \ldots, A^{j_i}_i$ be the connected components of $G[A_i]$. Furthermore, assume  that 
		$$\forall\ 1\le l\le j_i, \quad\frac{c(\delta(A_i^1))}{\pi(A_i^1)} \le \frac{c(\delta(A_i^{l})}{\pi(A_i^l)}.$$
		It follows that  (see proof of Proposition~\ref{pro:basicineq}(d))
		$\frac{c(\delta(A_i^1))}{\pi(A_i^1)} \le \frac{c(\delta(A_i)}{\pi(A_i)}$. 
			Define a partition $\mathcal{B'} = (B'_1, \ldots, B'_k)\in \mathcal{D}'_{k,0}(V(G))$ as follows. 
		$$ B'_1 \isdef  A_1 \cup \bigcup_{i=2}^{k} \bigcup_{l=2}^{j_i} A_i^l$$
		and 
		$$\forall 2\le j\le k, \quad B'_j \isdef A_j^1.$$
		
		Note that
		$$\forall 2\le i \le k, \quad  \pi(A_i) \le \pi(A_1) \le \pi(B'_1).$$ 
		
		For each vertex $v\in V(G)$, let $p_\mathcal{A}(v)$ and $p_\mathcal{B'}(v)$ denote the index of the part 
		containing $v$ in $\mathcal{A}$ and $\mathcal{B'}$, respectively. 
		Note that for each edge $e = uv$ with $u\in B'_1$ and $v\not\in B'_1$,  we have
		$$  \pi(B'_1) = \pi(B'_{p_\mathcal{B'}(u)}) \ge \pi(A_{p_\mathcal{A}(u)}).$$
		Thus
		\begin{equation}
			\label{eqn:b1bound}
			c(\delta(B'_1))/\pi(B'_1) = \sum_{e\in \delta(B'_1)} c(e)/\pi(B'_1) \le \sum_{e\in \delta(B'_1)} c(e)/\pi(A_{p_\mathcal{A}(e\cap B'_1) })
			\le  \Psi_G(\mathcal{A}).
		\end{equation}
		
		It also follows from the discussion above that 
		\begin{equation}
			\label{eqn:b2nbound}
			\sum_{i=2}^k c(\delta(B'_i))/\pi(B'_i) \le \sum_{i=2}^k c(\delta(A_i))/\pi(A_i) \le \Psi_G(\mathcal{A}). 
		\end{equation}
		
		Combining \ref{eqn:b1bound} and \ref{eqn:b2nbound} gives
		\[
		\Psi_G(\mathcal{B}) \le \Psi_G(\mathcal{B'}) = \sum_{i=1}^k c(\delta(B'_i))/\pi(B'_i) \le 2\Psi_G(\mathcal{A}) = 2\psi_{k,0}^1(G). 
		\]
		\item Next, we prove $ \frac{1}{2} \bar{\psi}_{k,0}^1(G) \le  \Psi_G(\mathcal{B})$. To see this, 
		Let $\mathcal{B} = \{B_1, \ldots, B_k\}$. We show that there is a way to merge each connected component of $G[B_1]$ with 
		one of its neighboring parts such that the resulting partition is at most twice as costly as $\mathcal{B}$. 
		
		Let the components of $G[B_1]$ be $D_1, \ldots, D_l$ and assume 
		$\frac{c(\delta(D_1))}{\pi(D_1)} \le \frac{c(\delta(B_{1})}{\pi(B_1)}$. 
		Also, for each $i\in\{2,\ldots, l\}$, let $N(D_i)$ denote the index of set of the parts neighboring $D_i$; i.e.\ 
		$j\in N(D_i)$ if and only if there is an edge between $B_j$ and $D_i$. 
		
			We construct a partition $\mathcal{A}=(A_1, \ldots, A_k)\in \mathcal{CD}_{k,0}$ from $\mathcal{B}$ through the following procedure. 
		Initially, set $A_1 := D_1$, and $A_i := B_i$ for each $2\le i \le k$. 
		
		\noindent	For $i=2,\ldots, l$, in turn:
		\begin{enumerate}
			\item Let $m := \mathrm{arg}\displaystyle{\max_{j\in N(D_i)}} \pi(B_j)$. 
			\item Set $A_m := A_m \cup D_i$.  
		\end{enumerate}
		
		Obviously each $A_i$ induces a connected subgraph, i.e.\ $\mathcal{A}\in \mathcal{CD}_{k,0}$. 
		Let $\delta(\mathcal{A})$ denote the set of edges appearing between any two sets of $\mathcal{A}$. 
		Observe that for any edge $\{u,v\}\in \delta(\mathcal{A})$ with $u\in B_1\setminus D_1$ and $v\not\in B_1$, the above procedure for constructing $A_j$'s
		guarantees that $\pi(A_{p_\mathcal{A}(u)}) \ge \pi(B_{p_\mathcal{B}(v)})$. 
		We are now ready to show that $\Psi_G(\mathcal{A}) \le 2\Psi_G(\mathcal{B})$.
		\renewcommand{\arraystretch}{2.5}
		{
			\[ {\everymath={\displaystyle}
				\begin{array}{rcl}
					\displaystyle
					\Psi_G(\mathcal{A}) &=&  \sum_{\ssc \{u,v\}\in \delta(\mathcal{A})} \left(\frac{c(\{u,v\})}{\pi(A_{p_\mathcal{A}(u)})}+\frac{c(\{u,v\})}{\pi(A_{p_\mathcal{A}(v)})}\right) \\
										&=&  \sum_{\substack{\ssc u\in B_1 \\ \ssc v\not\in B_1}} \left(\frac{c(\{u,v\})}{\pi(A_{p_\mathcal{A}(u)})}+\frac{c(\{u,v\})}{\pi(A_{p_\mathcal{A}(v)})}\right) %
										+ \sum_{\ssc \{u,v\} \cap B_1=\emptyset} \left(\frac{c(\{u,v\})}{\pi(A_{p_\mathcal{A}(u)})}+\frac{c(\{u,v\})}{\pi(A_{p_\mathcal{A}(v)})}\right) \\
										&=&  \frac{c(\delta(D_1))}{\pi(D_1)}  %
										+ \sum_{\substack{\ssc u\in B_1\setminus D_1 \\ \ssc v\not\in B_1}} \frac{c(\{u,v\})}{\pi(A_{p_\mathcal{A}(u)})} %
										+ \sum_{\substack{\ssc u\not\in B_1 \\ \ssc v\in V(G)}} \frac{c(\{u,v\})}{\pi(A_{p_\mathcal{A}(u)})} \\
										&\le&  \frac{c(\delta(B_1))}{\pi(B_1)}  %
										+ \sum_{\substack{\ssc u\in B_1\setminus D_1 \\ \ssc v\not\in B_1}} \frac{c(\{u,v\})}{\pi(B_{p_\mathcal{B}(v)})} %
										+ \sum_{\substack{\ssc u\not\in B_1 \\ \ssc v\in V(G)}} \frac{c(\{u,v\})}{\pi(B_{p_\mathcal{B}(u)})} \\ 
										&\le&  \sum_{\substack{\ssc u\in B_1 \\ \ssc v\not\in B_1}} \frac{c(\{u,v\})}{\pi(B_{p_\mathcal{B}(u)})} %
										+ 2\sum_{\substack{\ssc u\not\in B_1 \\ \ssc v\in V(G)}} \frac{c(\{u,v\})}{\pi(B_{p_\mathcal{B}(u)})} \\
										&\le& 2\Psi_G(\mathcal{B}). 
				\end{array}
			} \]
		}
	\end{itemize}

	This proves Part (a). For Part (b) note that
	\[
	\begin{array}{rcll}
		\bar{\psi}_{k,\rho}^1(G) &\le& \bar{\psi}_{k,0}^1(G) & \qquad \text{Proposition~\ref{pro:basicineq}, Part (b)} \\
		&\le& 4\psi_{k,0}^1(G) & \qquad \text{Part (a)} \\ 
		&\le& 8 \psi_{k,n}^1(G) & \qquad \text{Proposition~\ref{pro:basicineq}, Part (e)}\\
		&\le& 8 \psi_{k,\rho}^1(G). &  \qquad \text{Proposition~\ref{pro:basicineq}, Part (a)}.
	\end{array}
	\]

	
\end{proof}

Part (b) of Proposition~\ref{pro:connected} immediately yields

\begin{corollary}
	\label{cor:8alpha}
	A factor $\alpha$ polynomial approximation algorithm for $\mathsf{ISO}^1$ on a class of graphs yields a factor $8\alpha$ polynomial approximation algorithm for 
	$\mathsf{CRISO}^1$ on the same class of graphs.
\end{corollary}

By Corollaries~\ref{cor:2alpha}~and~\ref{cor:8alpha} in order to approximate the robust versions of the mean isoperimetry problem, it suffices to approximate 
$\mathsf{ISO}^1$, to which task we now turn in the remainder of this section. 

Before we proceed, let us recall the tree cut sparsifier introduced in \cite{racke2014improved} as follows.
\begin{alphthm}[\cite{racke2014improved}]
	\label{thm:Racke}
Given an edge-weighted graph $G=(V,E)$ on $n=|V|$ vertices and the weight function $c:E\to \mathbb{R}_+$, 
let $\lambda_G(U_1,U_2)$ denote the minimum weight of a subset of edges of $G$ whose removal 
disconnects the given subsets of vertices, $U_1$ and $U_2$. Then, there is a tree  $T=(V',E')$, where the leaves of $T$ correspond to the vertices of $G$, such that for any 
$U\subset V$, 
\[
	c(\delta_G(U))\le \lambda_T(U, V\setminus U) \le O(\log^{1.5}n\log\log n) c(\delta_G(U)).
\]
Moreover, if the edge-weights are polynomially bounded, then the tree $T$ may be constructed in polynomial time%
	\footnote{Although it is not explicitly mentioned in \cite{racke2014improved}, the polynomial-time construction of the tree cut-sparsifier needs a polynomial upper bound for the edge-weights as a consequence of Lemma~$2$ of the referenced article.}.
\end{alphthm}

Given a weighted graph $G = (V,E)$ with the weight  functions $c:E\to \mathbb{R}_+$ and $\pi:V\to \mathbb{R}_+$,
the modified R{\"a}cke-Shah tree cut sparsifier of $G$ is the weighted tree cut sparsifier introduced in Theorem~\ref{thm:Racke} equipped with the vertex-weight function which is one on each leaf and zero elsewhere.

\begin{theorem}
	\label{thm:hstapprox}
	If $T=(V',E')$ is the modified tree cut sparsifier of a weighted graph $G=(V,E)$ with weight functions $c:E\to \mathbb{R}_+$ and $\pi:V\to \mathbb{R}_+$, then
	\[
	\psi^1_k(G) \le  \ \psi^1_k(T) \le O(\log^{1.5}n \log\log n) \ \psi^1_k(G).
	\]
\end{theorem}

\begin{proof}
	Let $\mathcal{U'} = \{U'_1, U'_2, \ldots, U'_k\}$ and $\mathcal{S} = \{S_1, S_2, \ldots, S_k\}$ be subpartitions of $V'$ and $V$ minimizing 
$\psi^1_k(T)$ and  $\psi^1_k(G)$, respectively. We show 
	\[
	\Psi_G(\mathcal{S}) \le  \ \Psi_T(\mathcal{U'})\le O(\log^{1.5}n \log\log n) \ \Psi_G(\mathcal{S}).
	\]

To prove the first inequality, we first show that given $\mathcal{U'}$,
	there exists a subpartition $\mathcal{U} = \{U_1, U_2, \ldots, U_k\}$ of $V$ 
	for which we have,
	\[
	\Psi_G(\mathcal{U}) \le  \ \Psi_T(\mathcal{U'}).
	\]
	Let $\mathcal{U} = \{U_1, U_2, \ldots, U_k\}$ be the subpartition of $V$ defined as
	$U_i \isdef U'_i\cap V$
	(note that $U_i\ne \emptyset$ because $U'_i$ can not contain only zero-weight nodes). Then, for each $1\le i\le k$, we have 
	$$c(\delta_G(U_i)) \le \lambda_T(U_i, V\setminus U_i) \le c(\delta_T(U'_i)).$$
	Also, since $\pi(U_i) = \pi(U'_i)$, one may verify that 
	\[
		\Psi_G(\mathcal{U}) = \sum_{i=1}^{k}\frac{c(\delta_G(U_i))}{\pi(U_i)} %
		\le  \sum_{i=1}^{k}\frac{c(\delta_T(U'_i))}{\pi(U'_i)} \le  \ \Psi_T(\mathcal{U'}),
	\]
implying 
	\[
		\Psi_G(\mathcal{S}) \le  \ \Psi_G(\mathcal{U}) 
		\le \Psi_T(\mathcal{U'}).
	\]

	To prove the second inequality, we show that
	there exists a subpartition $\mathcal{S'} = \{S'_1, \ldots, S'_k\}$ of $V'$ such that 
	$$\Psi_T(\mathcal{S'}) \le O(\log^{1.5}n \log\log n) \ \Psi_G(\mathcal{S}).$$
	Without loss of generality, assume $w(S_1) \le w(S_2) \le \ldots \le w(S_k)$. 
	Also, define $S_{k+1} \isdef V\setminus \cup_{A\in \mathcal{S}} A$ to be the set of outliers and let  
    $V'_{i} \isdef V' \setminus \cup_{j<i} S'_j$ and $T_i \isdef T[V'_i]$. 
	Now, construct the subpartition  $\mathcal{S'} = \{S'_1, \ldots, S'_k\}$ inductively as follows:\\
	\ \\
	Choose $S'_i \subseteq V'_i$ such that 
	\begin{itemize}
		\item[{\rm 1.}] $S'_{i} \supseteq S_i $.
		\item[{\rm 2.}] $S'_i\cap S_j = \emptyset, \ \forall j > i$.
		\item[{\rm 3.}] The cut between $S'_{i}$ and $V'_{i+1}$ is minimum among all cuts separating $S_i$ from $\cup_{j>i} S_j$ in $T_i$.
	\end{itemize}
Then, we have 	
	\begin{equation}
		\label{eqn:cdelta}
		\begin{split}
			c(\delta_{T_i}(S'_i)) &= \lambda_{T_i}(S_i, \bigcup_{j>i} S_j) \le %
			\lambda_{T}(S_i, \bigcup_{j>i} S_j)\\
			& \le 
			\lambda_{T}(S_i, \bigcup_{j\ne i} S_j) \le O(\log^{1.5}n\log\log n) c(\delta_G(S_i)).
		\end{split}
	\end{equation}
	On the other hand,
	\[
		 \Psi_T(\mathcal{S'}) = 
		 \sum_{i=1}^{k} \frac{c(\delta_T(S'_i))}{\pi(S'_i)} = %
		 \sum_{i=1}^{k} \frac{(c(\delta_{T_i}(S'_i))+c(\delta_T(S'_i)\setminus\delta_{T_i}(S'_i)))}{\pi(S'_i)},
	\]
	and note that any edge $e\in (\delta_T(S'_i)\setminus\delta_{T_i}(S'_i))$ belongs to 
	$\delta_{T_j}(S'_j)$ for some 
	$j < i$, and hence, as the weights of $S_i$'s (and also $S'_i$'s) are nondecreasing, the contribution 
	of $e$ to $\sum_{i=1}^{k} \frac{c(\delta_{T_i}(S'_i))}{\pi(S'_i)}$
	which is $\frac{c(e)}{\pi(S'_j)}$ is at least as big as its contribution to 
	$\frac{c(\delta_T(S'_i)\setminus\delta_{T_i}(S'_i))}{\pi(S'_i)}$. Hence,
	\[
		\begin{split}
			\Psi_T(\mathcal{S'})& =  %
			\sum_{i=1}^{k} \frac{(c(\delta_{T_i}(S'_i))+c(\delta_T(S'_i)\setminus\delta_{T_i}(S'_i)))}{\pi(S'_i)}%
			\le 2 \sum_{i=1}^{k} \frac{c(\delta_{T_i}(S'_i))}{\pi(S'_i)}\\ 
			& \le  \sum_{i=1}^{k} O(\log^{1.5}n\log\log n) \frac{c(\delta_G(S_i))}{\pi(S_i)} = %
			O(\log^{1.5}n\log\log n)\Psi_G(\mathcal{S})
		\end{split}
	\]
	where that last inequality follows from \ref{eqn:cdelta}.  Therefore,
	\[
	\Psi_T(\mathcal{U'}) \le \Psi_T(\mathcal{S'})
		\le O(\log^{1.5}n \log\log n) \ \Psi_G(\mathcal{S}),
	\]
	where the first inequality holds since $\mathcal{\mathcal{U'}}$ is the optimal solution for $T$. 
\end{proof}

We now have all the necessary ingedients to prove the main result of this section. 

\begin{theorem}\label{thm:mainapprox}
	There is a polynomial $O(\log^{1.5}n \log\log n)$-approximation algorithm for $\mathsf{ISO}^{1}$, $\mathsf{CISO}^{1}$,
	$\mathsf{RISO}^{1}$, and $\mathsf{CRISO}^{1}$ on weighted graphs $G$ with polynomially bounded weights. 
\end{theorem}

\begin{proof}
	The solvability of $\mathsf{ISO}^{1}$ with the specified approximation factor follows from Theorems~\ref{thm:hstapprox} and~\ref{thm:tree}. 

	The conclusion for $\mathsf{RISO}^{1}$ and $\mathsf{CRISO}^{1}$ follows from Corollaries~\ref{cor:2alpha} and~\ref{cor:8alpha}, and 
	the conclusion for $\mathsf{ISO}^{1}$. 

	Theorem~\ref{thm:daneshgarjavadi-connected} and the conclusion for $\mathsf{ISO}^{1}$ yield the approximability of $\mathsf{CISO}^{1}$. 
\end{proof}

\begin{remark}
	\label{rem:risooptimizer}
	The proof of Theorem~\ref{thm:hstapprox} combined with 
	Remark~\ref{rem:isooptimizer}, implies that one may also find the actual optimizing subpartition with the prescribed approximation ratio for 
	$\mathsf{ISO}^1(G,k)$ in polynomial time. 
	Likewise, as the proofs of Propositions~\ref{pro:basicineq}(e)\footnote{See \cite{daneshgar-javadi-trees}}%
	 and \ref{pro:connected} are constructive,  the same is true for finding the approximately optimal subpartitions for 
	 $\mathsf{RISO}^1(G,k,\rho)$ and $\mathsf{CRISO}^1(G,k,\rho)$. 
\end{remark}

\section{Acknowledgements}
\label{sec:acknowledgement}
The authors would like to thank Arash Ahadi for making useful suggestions regarding the presentation of the algorithm in Section~\ref{sec:trees}, and   
Ramin Javadi and Saleh Ashkboos for suggesting the modification to the R{\"a}cke-Shah tree cut sparsifier used in Section~\ref{sec:general}.



\end{document}